\documentclass[12pt,a4paper]{article}
\usepackage{geometry}
\usepackage[small]{titlesec}
\usepackage{xcolor} 
\usepackage{hyperref}
\usepackage[utf8]{inputenc}
\usepackage{amsmath}
\usepackage{mathtools,amssymb}
\usepackage{amsfonts}
\usepackage{amssymb}
\usepackage{physics}
\usepackage[autostyle]{csquotes}
\usepackage{graphicx}
\graphicspath{{./figures/figure_outline/}}
\usepackage{fancyhdr}
\usepackage{cite}
\usepackage[noblocks]{authblk}
\usepackage[switch, modulo]{lineno}
\usepackage[margin=1cm]{caption}
\hypersetup{
colorlinks=true,
linkcolor=blue,
citecolor=magenta,
filecolor=magenta,
urlcolor=cyan
}
\begin{document}
\title{A hybrid full-wave Markov chain approach to calculating radio-frequency wave scattering from Scrape-off Layer filaments}

\author[1]{Bodhi Biswas\thanks{E-mail: \texttt{bodhib@mit.edu}}}
\author[2]{Syun'ichi Shiraiwa}
\author[1]{Seung-Gyou Baek}
\author[1]{Paul Bonoli}
\author[1]{Abhay Ram}
\author[1]{Anne White}
\affil[1]{Plasma Science and Fusion Center, Massachusetts Institute of Technology}
\affil[2]{Princeton Plasma Physics Laboratory}

\maketitle
\begin{abstract}
	The interaction of radio-frequency (RF) waves with edge turbulence modifies the incident wave-spectrum, and can significantly affect RF heating and current drive in tokamaks. Previous LH scattering models have either used the weak-turbulence approximation, or treated more realistic, filamentary turbulence in the ray-tracing limit. In this work, a new model is introduced which retains full-wave effects of RF scattering in filamentary turbulence. First, a Mie-scattering technique models the interaction of an incident wave with a single Gaussian filament. Next, an effective differential scattering-width is derived for a statistical ensemble of filaments. Lastly, a Markov chain solves for the transmitted wave-spectrum in slab geometry. This model is applied to LH launch for current drive. The resulting wave-spectrum is asymmetrically broadened in wave-number angle-space. This asymmetry is not accounted for in previous LH scattering models. The modified wave-spectrum is coupled to a ray-tracing/Fokker-Planck solver (GENRAY/CQL3D) to study its impact on current drive. The resulting current profile is greatly altered, and there is significant increase in on-axis current and decrease in off-axis peaks. This is attributed to a portion of the modified wave-spectrum that strongly damps on-axis during first-pass.
	
\end{abstract}
\bibliographystyle{iopart-num}
\newpage
\tableofcontents
\newpage
\section{Introduction}
\label{sec:1}
Before an external RF wave can damp in the core plasma of a magnetic confinement device, it must first propagate through the highly turbulent scrape-off layer (SOL) region. SOL turbulence is comprised of dense, coherent structures called blobs/filaments\cite{zweben2002edge,kirk2006filament} that can significantly modify the incident wave-spectrum. Scattering from filaments leads to refraction of the intended wave-path and broadening of the incident wave-spectrum, which in turn can cause lower efficiency in the intended function of the wave. For example, simulations predict significant power-loss through filament-assisted mode-conversion for launched ion-cyclotron waves\cite{tierens2020filament}. Electron cyclotron beams can be broadened in the presence of SOL turbulence, leading to ineffective targeting of neo-classical tearing modes \cite{tsironis2009electron}. In the case of driving current using Lower Hyrbid (LH) waves, SOL scattering is a promising explanation for the spectral gap problem and current drive density limit.

Measurements on Alcator C-Mod\cite{mumgaard2015lower}, EAST\cite{ding2018review}, and Tore Supra\cite{peysson2000progress} indicate self-similar, on-axis peaked LH current profiles. Ray-tracing/Fokker-Planck simulations predict off-axis peaks\cite{mumgaard2015lower}, which are inconsistent with these measurements. In addition, these simulated profiles are sensitive to plasma and wave launch parameters, unlike experiment. Lastly, Lower Hybrid current drive (LHCD) suffers from an anomalous density limit, beyond which current drive (CD) efficiency dramatically falls \cite{wallace2010absorption}. Meanwhile, SOL turbulence increases with Greenwald density\cite{cziegler2010experimental}. Raising the Ohmic current, and therefore decreasing Greenwald density and shrinking the SOL width is shown to increase LHCD efficiency at high densities in C-Mod\cite{baek2018observation}. These considerations suggest there are important spectral broadening effects, i.e. scattering from SOL turbulence, unaccounted for in the standard ray-tracing/Fokker-Planck model.

It should be noted that alternate spectral broadening mechanisms exist. These include full-wave effects in the core like interference and focusing, and edge mechanisms such as parametric decay instabilities (PDI)\cite{porkolab1977parametric}. The capability to run full-wave simulations of LHCD is fairly recent\cite{wright2009assessment, shiraiwa2010plasma}, and it is not yet clear whether it provides a better match to experiment than ray-tracing/Fokker-Planck models. PDI is a strong candidate for explaining the current drive density limit \cite{cesario2014spectral, baek2015high}. However, there is no clear indication that PDI significantly modifies the wave-spectrum in low-density discharges\cite{baek2015high}. Note that the parallel wave-vector up-shift from PDI and the perpendicular wave-vector rotation from scattering may both be required to bridge the LH spectral gap \cite{biswas2020study}. (The terms ``perpendicular/parallel'' are used in relation to the local background magnetic field.)

Early attempts to model LH wave scattering in the SOL treat the turbulence as incoherent drift-wave-like density fluctuations\cite{bellan1978effect,bonoli1982toroidal,andrews1983scattering}. This results in a diffusive process leading to the angular broadening of the perpendicular wave-vector component $\bold{k}_{\perp}$. While these models can significantly broaden the incident wave-spectrum, they have been unable to explain experimental measurements at either low or high densities\cite{peysson2011rf,bertelli2013effects}. A recent study models LH scattering from coherent SOL filaments with ray-tracing \cite{biswas2020study}. This results in an increased angle-broadening effect compared to previous models, which in turn leads to relatively better match with experimental current drive measurement and reduced sensitively to simulation parameters. Another recent study using full-wave simulations predict large parasitic loss of LH power in the presence of SOL filaments\cite{lau2020full}. This is attributed to significant partial-reflection and side-scattering. In ray-tracing, partial-reflection is neglected, and side-scattering is likely underestimated.

These results motivate a closer study of wave scattering from filaments using a full-wave treatment. Unlike ray-tracing, which only accounts for refraction and total-reflection, a full-wave model also retains the physical optics effects of interference, diffraction, and focusing. In addition, full-wave models can account for asymmetric scattering, resulting in the rotation of $\bold{k_{\perp}}$ in one preferential direction. Notably, this effect is ignored in ray-tracing and other wave-kinetic models for LH wave scattering. This is discussed further in \hyperref[sec:7.2]{Section 7.2}.

In this paper, a hybrid method is introduced to efficiently calculate the full-wave effects of RF scattering through a slab layer comprised of filaments. First, the scattered EM wave is calculated for an incident wave interacting with a single filament, which is modeled as an infinitely long cylinder. This problem has a semi-analytic solution, and can be very efficiently computed relative to numeric full-wave solvers. Previous implementations\cite{myra2010scattering,ram2016scattering} of this semi-analytic scattering (SAS) model in a plasma-physics context were restricted to ``flat-top'' (homogenuous) filaments. This model is generalized to filaments with radially-varying density profiles, which better mimic experimentally relevant filaments. Next, a scattering-width (analagous to a scattering cross-section) is calculated from the scattered wave solution.  Third, this process is repeated multiple times, for different filament parameters, until a statistically averaged ``effective'' scattering-width is produced. Lastly, this effective scattering-width is used to calculate the cumulative effect of multiple scattering events for an RF wave incident on a turbulent slab. This is identical to solving the \emph{radiative transfer equation}, for which many techniques exist from the fields of optics and neutronics. The present study uses the absorbing Markov chain technique\cite{esposito1978radiative} to compute the final transmitted and reflected wave-spectrum. This work-flow is henceforth called the Semi-Analytic Scattering Markov Chain (SAS-MC) model.

The SAS-MC model can be applied to any frequency range to study RF-scattering in the SOL because it is derived using the fully-electromagnetic cold dispersion relation. This paper focuses on applying it to LH waves. Assuming certain properties about the SOL geometry and turbulence, a modified wave-spectrum is calculated for LH launch in a low-density Alcator C-Mod discharge. This wave-spectrum is coupled to the ray-tracing/Fokker-Planck solver GENRAY\cite{smirnov2001genray}/CQL3D\cite{harvey1992cql3d} to determine its impact on current drive. The result is a significantly modified CD profile that is peaked on-axis. This increased on-axis damping is attributed to a fraction of LH rays rotated by scattering such that they damp on-axis during first-pass. In addition, a mechanism for asymmetric scatter is identified. The extent of asymmetric scattering increases with background density and turbulence.

This paper is structured in the following way. \hyperref[sec:2]{Section 2} reviews the SAS model for calculating the scattered wave. In \hyperref[sec:3]{Section 3}, the SAS model is generalized to radially-varying filaments. \hyperref[sec:4]{Section 4} discusses the calculation of the scattering-width and ``effective" scattering width. \hyperref[sec:5]{Section 5} introduces the Markov chain (MC) model necessary to calculated the final modified wave-spectrum following propagation through the SOL. In \hyperref[sec:6]{Section 6}, the SAS-MC model is compared to the higher fidelity numeric full-wave solver PETRA-M\cite{shiraiwa2017rf}. Limitations to the accuracy of the SAS-MC model are discussed. \hyperref[sec:7]{Section 7} applies the SAS-MC model to Lower Hybrid launch in a typical SOL in C-Mod. There is an in-depth discussion about the asymmetric profile of the scattering-width.  Comparisons are made with a ray-tracing treatment. In \hyperref[sec:8]{Section 8}, the modified wave-spectrum is coupled to GENRAY/CQL3D to model LHCD in a C-Mod discharge. \hyperref[sec:9]{Section 9} summarizes the results of this study.

\section{Review of Semi-analytic scattering model}
\label{sec:2}
The first component of the SAS-MC model is the semi-analytic Mie-scattering description of an incident RF wave interacting with a single cylindrical filament. To the authors' knowledge, this problem is first treated in the magnetized plasma context by Myra \& D'Ippolito (2010)\cite{myra2010scattering} in the Lower Hybrid limit ($\Omega_{ci}^2 \ll \omega^2 \ll \Omega_{ce}^2$) for a homogenous cylinder. Ram \& Hizanidis (2016)\cite{ram2016scattering} extended this model to all frequencies. The single filament scattering model is briefly reviewed in this section, and then extended to radially in-homogeneous filaments in \hyperref[sec:3]{Section 3}.

\begin{figure}[!h]
\centering
\includegraphics[width=8cm, height=7cm]{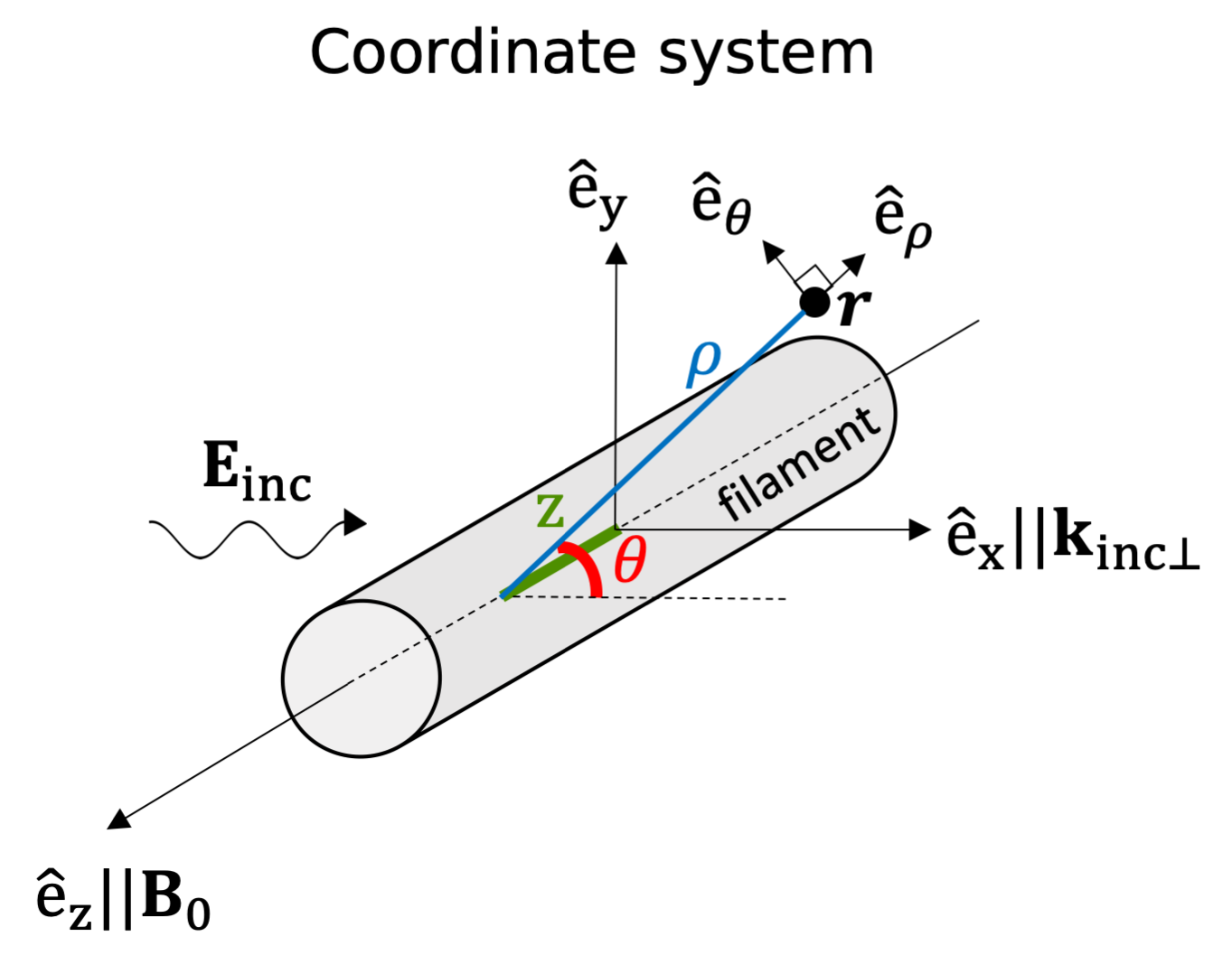}
\caption[font=5]{Poloidal and Cartesian coordinate system used to model RF scattering from a field-aligned filament.}
\label{fig:1}
\end{figure}

Fig. \ref{fig:1} illustrates the SAS model coordinate system. The cylinder axis is aligned with the background magnetic field $\bold{B}_0$ in the z-direction. Given a plane-wave traveling in the +x-direction, the objective is to calculate the scattered wave exterior to the cylinder. The electric field inside and outside the filament must satisfy the vector wave equation. In the case that the cylinder's dielectric properties have no longitudinal ($z$) and poloidal ($\theta$) dependence, this problem can be solved via separation of variables in cylindrical coordinates.

Since the medium is homogeneous inside and outside the cylinder, it is simple to formulate an ansatz to the wave equation in each region. There are five waves to consider: the known incident wave $\bold{E}_{0}$; the scattered slow, fast wave $\bold{E}_{1}$, $\bold{E}_{2}$ outside the cylinder; and the slow, fast wave $\bold{E}_{3}$, $\bold{E}_{4}$ excited inside the cylinder. At the discontinuous boundary $\rho = a_b$, the fields inside and outside must satisfy Maxwell's boundary conditions.

\subsection{Ansatz to electric field}
\label{sec:2.1}
Using separation of variables (see \hyperref[sec:A1]{App. A}), the electric field can be written in cylindrical coordinates as:
\begin{subequations}
\begin{align}
E_{j \gamma} & = e^{i(k_{||}z-\omega t)}\sum_{m=-\infty}^{+\infty} E_{jm} W_{j \gamma m} e^{i m \theta}; \quad \quad\gamma = \rho,\theta,z \\
W_{j\rho m} & = \xi_{jx} J_{m}^{'}(k_{j \perp}\rho) - i\xi_{jy} \frac{m}{k_{j \perp} \rho} J_{m}(k_{j \perp}\rho) \\
W_{j\theta m} & = i\xi_{jx} \frac{m}{k_{j \perp} \rho} J_{m}(k_{j \perp}\rho) + \xi_{jy} J_{m}^{'}(k_{j \perp}\rho) \\
W_{jzm} & = i \xi_{jz}J_{m}(k_{j \perp}\rho)
\end{align}
\end{subequations}
where $j=0,...,4$ is the wave index. $\bar{\xi}_j = \{\xi_{jx},\xi_{jy},\xi_{jz}\}$ is the plane-wave polarization of wave $j$. $J_{m}$ is the Bessel function of the first kind and order $m$. $J_{m}^{'}$ is the first derivative of $J_{m}$ with respect to its argument. For the known incident plane wave ($j=0$), it is required that:
\begin{equation}
E_{0m} = i^{m-1}
\end{equation}
$E_{jm}$ for $j>0$ have yet to be determined.

Note that $k_{||}$ is the same for all waves, and is fixed by the incident wave. This is a consequence of Snell's law applied to a medium that is constant along the z-direction.

\subsection{Boundary conditions}
\label{sec:2.2}
In general, the solutions in equations (1) can have both $J$ and $Y$ terms, where $Y$ is the Bessel function of the second kind. The requirement that $\bold{E}$ is finite at $\rho = 0$ leads to $Y$ terms being zero for the slow and fast branch inside the filament. For $\rho \rightarrow \infty$, the scattered fields must be radiating away from the filament. For the scattered fast wave, this requires the use of Hankel functions of the first kind, $H^{1}$, instead of $J$ in eqs. (1). The LH slow wave is backward-propagating, meaning $\bold{k_{\perp}}$ and $\bold{v}_{gr,\perp}$ are anti-parallel. This requires the use of Hankel functions of the second kind, $H^{2}$, for the scattered slow wave.

It should be noted that a backward-propagating incident wave (ie. the slow wave) has $k_{inc,x} = -k_{\perp}$. This flipped sign can most easily be accounted for by substituting $J_{m} \rightarrow J_{-m}$\cite{myra2010scattering}.

Lastly, a system of equations must be formulated to determine coefficients $E_{jm}$ for $j=1,...,4$. This is accomplished by imposing the four independent Maxwell boundary conditions at $\rho = a_b$ (see \hyperref[sec:A2]{App. B}). For each poloidal mode-number, there are four unknown coefficients and four boundary conditions, resulting in a solvable system of equations.

\section{Generalizing to radially in-homogeneous filaments}
\label{sec:3}
The scattering model reviewed in the previous section is now extended to account for radially in-homogeneous cylinders. The cylinder remains poloidally symmetric, and therefore the poloidal mode-numbers are still uncoupled. A solution via separation of variables, similar to that in \hyperref[sec:2]{Section 2}, is still possible. The following solution scheme for a radially in-homogeneous filament is similar to Mie-scattering formulations for scattering from layered dielectrics \cite{kai1995finely} or annular cylinders\cite{wu1994scattering}. To the authors' knowledge, this is its first application in the context of Lower Hybrid wave scattering.

In the previous case of a totally homogeneous ``flat-top" cylinder, there was a single boundary (at $\rho = a_b$) and therefore only one radial ``bin'' inside the cylinder. The cylinder is now discretized into multiple bins $r=0,...,R$. In other words, the filament is now a set of radially-stratified concentric cylinders. This introduces discontinuities in the media between bins, and so boundary conditions must be imposed at each separating layer. In the limit $R \rightarrow \infty$, a cylinder with a smoothly varying radial profile can be modeled with arbitrary precision.
\subsection{Modified system of equations}
\label{sec:3.1}
\begin{figure}[!h]
\centering
\includegraphics[width=11cm, height=6cm]{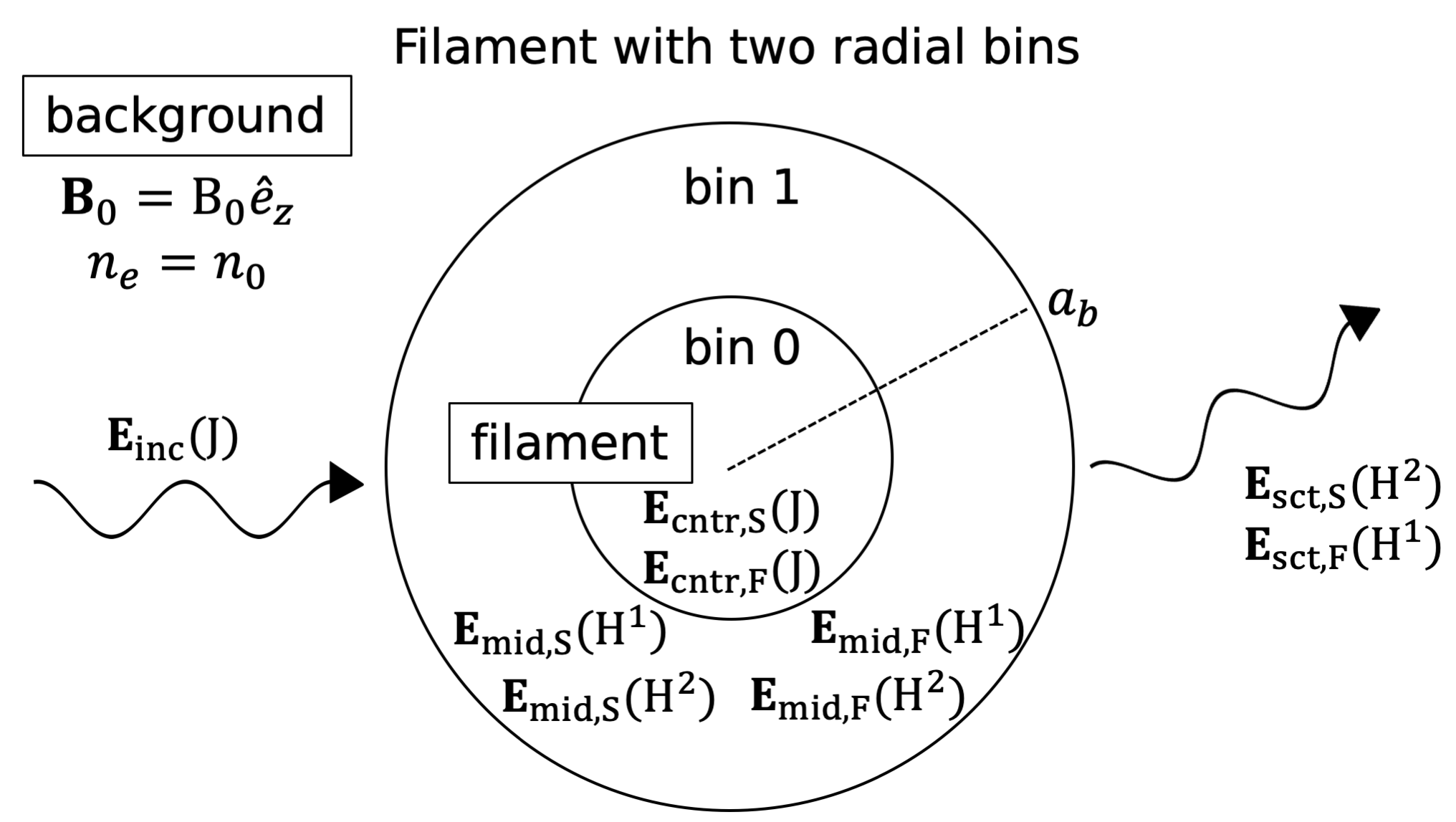}
\caption[font=5]{The different contributions of $\bold{E}$ in the SAS model for a filament with two radial bins ($R=1$). There are two regions of interest: background and filament. The filament is further divided into the central (cntr) and intermediate (mid) region. `S' and `F' denote slow and fast LH branches. Functions in parentheses denote the type of Bessel function used in eq. (1).}
\label{fig:1b}
\end{figure}

Remember that the ``flat-top'' ($R=0$) system is solvable because there exist four unknowns $E_1,E_2,E_3,E_4$ and four independent boundary equations for each mode-number $m$. A similar system of equations must be derived for the general $R>0$ case. The simplest case ($R=1$) is illustrated in Fig \ref{fig:1b}. In the intermediate layers ($0<r \leq R$), each wave branch is generally a function of both $H_{m}^{1}$ and $H_{m}^{2}$ terms, and are therefore split into these two electric field contributions. 

With the considerations made above, there are now eight unknown waves for the $R=1$ case. There are also two boundaries, supplying four boundary conditions each. This results in a solvable system for the total electric field everywhere.

For the general case ($R>0$), there are all together $4(R+1)$ equations for each mode-number. For more detail, see \hyperref[sec:A2]{App. C}. Solving for the total electric field requires inverting $4(R+1) \times 4(R+1)$ matrices a total of $2M+1$ times, where $M$ is the maximum mode-number chosen to truncate the series. These matrices are sparse and banded, resulting in fast solution times on the order of seconds on a single CPU.

\subsection{Poloidal and radial resolution}
\label{sec:3.2}
The electric field will evolve on three possible length-scales: $a_b, k_{\text{in} \perp}^{-1}$, or $k_{\text{out} \perp}^{-1}$. Sub-scripts ``in'', ``out'' denote inside, outside the cylinder. Define a characteristic poloidal mode-number $\tilde{m} \equiv \text{max}(k_{\text{in} \perp},k_{\text{out} \perp})a_{b}$. If $\tilde{m} \gtrsim 1$, terms with $|m|>\tilde{m}$ will rapidly decay in magnitude. Therefore, for a converged solution, it is necessary that $M \gg \tilde{m}$. If $\tilde{m}\ll 1$, it is necessary that $M \gg 1$.

A rule-of-thumb can also be derived for how many radial bins are required. The source of error stems from discretizing the cylinder's smoothly varying radial profile into homogeneous radial bins. The discontinuity between bins is of order $L_{n}^{-1} \Delta r$, where $L_n$ is the characteristic length of the density in-homogeneity.  $\Delta r$ is the radial bin width. Assume the cylinder has a monotonically decreasing radial profile with characteristic radial width $a_b$. Then the discontinuity is approximately $L_{n}^{-1} \Delta r \sim \left( \frac{n_b}{n_0 a_b}\right) \left( \frac{a_b}{R}\right) = \frac{n_b}{n_0 R}$. $\frac{n_b}{n_0}$ is the ratio of peak (center) cylinder density to the background density, and $R$ is the number of radial bins. To ensure the discontinuities are small requires $R \gg \frac{n_b}{n_0}$.

\section{Scattering-width for statistical ensemble of filaments}
\label{sec:4}
While the model described above solves for the scattered field, it is more convenient to calculate a \emph{differential scattering-width}, which accounts for the deflection of scattered power in $\theta$-space. The differential scattering-width is defined as \cite{myra2010scattering}:
\begin{equation}
\sigma(\theta) \equiv \frac{d\sigma}{d\theta} = \frac{\textrm{lim}_{\rho \rightarrow \infty} \rho S_{sct,\rho}(\rho,\theta)}{S_{inc,x}}
\end{equation}
 $\bold{S}_{inc}$, $\bold{S}_{sct}$ is the incident, scattered Poynting flux. It follows that the \emph{scattering-width} $\sigma = \int_{-\pi}^{+\pi} \sigma (\theta) d\theta$. See \hyperref[sec:A4]{App. D} for derivation and physical meaning of $\sigma$. The scattering-width (units of length) is the 1D analogy for a scattering cross-section (units of length squared).

Lastly, define a \emph{normalized differential scattering-width} $\hat{\sigma}(\theta) = \sigma(\theta)/\sigma$ which will be useful in \hyperref[sec:5]{Section 5}. Note that $\sigma(\theta) \geq 0$ since, at a far enough distance, the scattered waves must be radiating $away$ from the cylinder.

So far, no explicit expression for $S_{sct,\rho}$ and $S_{inc,x}$ in eq. (3) have been provided. The time-averaged Poynting flux for the scattered wave can be written as:
\begin{equation}
\bold{S}_{sct,j}= \frac{1}{2 \mu_0 \omega} \Im{\bold{E}_{j}^{*} \cross (\nabla \cross \bold{E}_{j})}
\end{equation}
where $j = S,F$ now refer to the scattered slow and fast wave, respectively.
Equations (1) are substituted into eq. (4), and then substituted into eq. (3) to produce
\begin{equation}
\sigma_{j}(\theta) = \frac{ \mp 2}{\pi} \frac{|\xi_{j,y}|^2 + |\xi_{j,z}|^2 + \frac{k_z}{k_{j\perp}} \Re{\xi_{j,x} \xi_{j,z}^{*}} }{k_{inc,x} (|\xi_{inc,y}|^2 + |\xi_{inc,z}|^2) -k_{z} \Re{\xi_{inc,x} \xi_{inc,z}^{*}}} \abs{\sum_{m=-\infty}^{+\infty}i^{\pm m} E_{jm} e^{i m \theta}}^{2}
\end{equation}
assuming real valued $k_{j\perp}$. If $k_{j\perp}$ is imaginary, then the RHS of eq. (5) is zero.
$\sigma_{(S,F)} (\theta)$ denotes the differential scattering-width for coupling from the incident slow wave to a scattered slow, fast wave. The asympotic relation $H_{m}^{(1,2)}(\tau)\approx \sqrt{\frac{2}{\pi \tau}}e^{\pm i(\tau-m \pi /2 - \pi/4)}$ for large argument $\tau$ has been used. In the denominator, $\bar{\xi}_{inc}$ is the normalized polarization of the incident slow-wave. Equation (5) is a generalization of the $\sigma (\theta)$ calculated by Myra \& D'Ippolito (2010)\cite{myra2010scattering}, which was done in the electrostatic limit. Equation (5) accounts for a fully electromagnetic dispersion tensor, and is therefore valid for the low densities in the far-SOL.

The total differential scattering-width is:
\begin{equation}
\sigma(\theta) =  \sigma_{S}(\theta) + \sigma_{F}(\theta)
\end{equation}
For background densities in which the slow-wave is propagating but the fast-wave is evanescent, $\sigma_{F} = 0$. It only becomes comparable to $\sigma_S$ as the background density approaches the mode-conversion density. For the purposes of studying slow-wave scattering in the SOL, it is reasonable to neglect $\sigma_{F}$.

\subsection{Effective differential scattering-width: $\sigma_{\text{eff}}(\theta)$}
\label{sec:4.1}
In the presence of a filament, the resulting scattered field will depend on $a_b$ and $n_{b}$. Therefore, $\sigma(\theta)= \sigma(\theta; n_b/n_0, a_b)$ for a given $n_0, \bold{B},N_{||}$, and $\omega$. A joint probability distribution function (PDF), $p(n_b/n_0,a_b)$, is introduced. This is the probability that a filament will have certain parameters $a_b$ and $n_b/n_0$. Taking a weighted-average of $\sigma(\theta)$ with this joint-PDF returns the statistically averaged $\sigma(\theta)$ for scattering from a randomly selected filament. This averaged, or \emph{``effective'' differential scattering-width} is defined as:
\begin{equation}
\sigma_{\text{eff}}(\theta) = \int_{0}^{\infty} da_{b} \int_{0}^{\infty}d \left(n_{b}/n_0\right) \, \sigma_{S}(\theta; n_b/n_0, a_b) p(n_b/n_0,a_b)
\end{equation}
where $p(n_b/n_0,a_b)$ is normalized such that $\int_{0}^{\infty} da_{b} \int_{0}^{\infty}d\left(n_{b}/n_0\right)p(n_b/n_0,a_b) = 1$. Again, $\sigma_{F}$ is neglected since the focus is on SOL plasmas.

The choice of joint-PDF for the filament parameters is guided by experimental measurements. In the SOL of C-Mod, mirror Langmuir probe measurements reveal positively skewed PDFs of density fluctuations\cite{graves2005self}. SOL fluid codes predict that filament width and density are positively correlated \cite{decristoforo2020blob}. Therefore, the joint-PDF is reasonably well described by a positively-skewed Gaussian PDF for each parameter ($a_b$ and $n_b/n_0$) along with a positive bi-variate correlation. The mean filament width and relative density is
\begin{subequations}
\begin{align}
\langle a_b \rangle & = \int_{0}^{\infty} da_{b} \int_{0}^{\infty}d \left(n_{b}/n_0\right) \, \, a_b \, p(n_b/n_0,a_b)\\
\left\langle  \frac{n_b}{n_0}\right\rangle & = \int_{0}^{\infty} da_{b} \int_{0}^{\infty}d \left(n_{b}/n_0\right) \, \, \frac{n_b}{n_0} \, p(n_b/n_0,a_b)\\
\end{align}
\end{subequations}
Filament mean width $\langle a_b \rangle$ is well bounded by gas-puff imaging (GPI) measurements as well as theory/simulation\cite{zweben2002edge, krasheninnikov2008recent, keramidas2020comparison}. Langmuir probe and GPI measurements provide a rough lower-bound on the filament mean relative density $\langle n_b/n_0 \rangle$, though this value will vary significantly at different radial locations in the SOL\cite{terry2003observations, zweben2002edge}.

\section{The radiative transfer equation in slab geometry}
\label{sec:5}
The previous sections deal with a single scattering event due to one filament. \hyperref[sec:4.1]{Section 4.1} introduced an effective scattering-width $\sigma_{\text{eff}}(\theta)$, but this still only gives information about the average scattered power due to \emph{one} filament. Consider a turbulent medium with filaments of mean width $\langle a_b \rangle$ with packing fraction $f_{p}$.  An incident LH wave will, on average, interact with  $\frac{f_p L_x}{\pi \langle a_{b} \rangle^{2}}$ filaments per unit length in the perpendicular plane. Therefore, $\Sigma_{\text{eff}} \equiv \frac{f_p}{\pi \langle a_b \rangle^{2}}\sigma_{\text{eff}}$ is the inverse mean-free-path for the incident power to scatter. A \emph{radiative transfer equation} (RTE) can then be derived.
\begin{equation}
\left(\frac{\partial}{\partial t} + \bold{v}_{gr} \cdot \nabla \right) P(\bold{r},\theta)= -\Sigma_{\text{eff}} |\bold{v}_{gr \perp}| P(\bold{r},\theta) + \Sigma_{\text{eff}} |\bold{v}_{gr \perp}| \int_{-\pi}^{\pi}\hat{\sigma}_{\text{eff}}(\theta-\theta^{'})P(\bold{r},\theta^{'})d\theta^{'}
\end{equation}
where $P(\bold{r},\theta)$ is the power density at $\bold{r}$ directed along angle $\theta$. In eqn. (9), the first term on the RHS accounts for power directed along $\theta$ that is lost to scattering. The second term accounts for power gained at $\theta$ due to scattering from all other $\theta^{'}$. (Any losses due to the anti-Hermitian part of the dielectric tensor are ignored.)

Now, consider a steady-state slab geometry with a filamentary turbulent layer limited to $0<x<L_{x}$. The filaments and background B are aligned along the z-direction. A LH plane-wave is incident on the slab from the left ($S_{inc,x}>0$). Since the background density is homogeneous, $|\bold{v}_{gr\perp}|$ is constant.  Equation (9) then simplifies to

\begin{subequations}
\begin{align}
& cos\theta \frac{dP(x,\theta)}{dx} = -\Sigma_{\text{eff}}P(x,\theta) + \Sigma_{\text{eff}}\int_{-\pi}^{\pi}\hat{\sigma}_{\text{eff}}(\theta-\theta^{'})P(x,\theta^{'})d\theta^{'}\\
& P(L_{x},\theta) = 0 \quad \textrm{for} \quad |\theta| \geq \pi/2\\
& P(0,\theta) = \delta(\theta) \quad \textrm{for} \quad |\theta| \leq \pi/2
\end{align}
\end{subequations}
where $\delta (\theta)$ is the Dirac delta function. Compare this to eq. (31) in Andrews \& Perkins (1983)\cite{andrews1983scattering}, where a similar RTE is formulated for drift-wave-like turbulence. Equations (10b \& c) enforce no scattering back into the turbulent layer at $x = L_{x}$ and $0$, respectively.  Equation (10c) also enforces a normalized incident power from the left. Solving this equation and evaluating $P(x,\theta)$ at $x = L_x$ and $0$ results in the normalized angle-broadened transmitted and reflected wave-spectrum, respectively.

It should be noted that two critical assumptions have been made in formulating the RTE. (1) $\sigma_{\text{eff}} (\theta)$ is formulated using the far-field limit. (2) The interaction of a wave with multiple filaments is modeled by chaining multiple single-filament scattering events. Together, they constute the \emph{far-field approximation}, which is only valid if $\langle a_b \rangle \ll d$, and $k_{\perp} d \gg 1$, where $d$ is the average distance between filaments \cite{mishchenko2014electromagnetic}. This approximation breaks down as $f_p$ increases (and therefore $d$ decreases), and is further discussed in \hyperref[sec:6.2]{Section 6.2}. 

\subsection{Solution to RTE using a Markov chain}
\label{sec:5.1}
Equation (10) is an integro-differential equation, and cannot, in general, be solved analytically. One numerical method is to discretize the wave-spectrum into photons/rays, and stochastically evolve their trajectories, as per the standard Monte-Carlo technique. This method is rather slow in providing a converged wave-spectrum near $\theta = \pm \pi/2$, where tally counts are usually low. Given how simple the slab geometry is, a more elegant Absorbing Markov chain method can be employed. This Markov chain (MC) method is deterministic, so it avoids the low tally count problem. It is commonly used to solve for a reflected and transmitted wave-spectrum through a turbid slab (e.g. solar rays interacting with Earth's atmosphere). The present study closely follows the formalism by Esposito \& House (1977)\cite{esposito1978radiative} and Xu \emph{et al.} (2011)\cite{xu2011markov}.

Power is incident on the turbulent slab from the left, directed along the x-direction ($\theta_{0}=0$). It is convenient to define a $\zeta \equiv |\cos{\theta}|$ such that $\zeta_{0}=1$. It is simple to calculate the fraction of transmitted power that does not scatter in the slab. This is the ``ballistic'' fraction:
\begin{equation}
P_{ball} = e^{-\frac{L_x}{\zeta_{0}}\Sigma_{\text{eff}}}
\end{equation}
It is also straightforward to calculate the transmitted and reflected fraction that only scatter \emph{once} in the slab:
\begin{equation}
P_{T,1}(\theta) = \hat{\sigma}(\theta-\theta_{0})e^{-\frac{L_{x}}{\zeta}\Sigma_{\text{eff}}}\times \begin{cases}
0 & \cos{\theta} < 0\\
\frac{\zeta \zeta_{0}}{\zeta-\zeta_{0}}\left[1-e^{-L_{x}\Sigma_{\text{eff}}(\frac{1}{\zeta_0}-\frac{1}{\zeta})}\right] & \theta \neq \theta_{0}\\
\Sigma_{\text{eff}}L_{x} & \theta = \theta_0
\end{cases}
\end{equation}
\begin{equation}
P_{R,1}(\theta) = \hat{\sigma}(\theta-\theta_{0}) \times \begin{cases}
0 & \cos{\theta} > 0\\
\frac{\zeta \zeta_{0}}{\zeta+\zeta_{0}}\left[1-e^{-L_{x}\Sigma_{\text{eff}}(\frac{1}{\zeta_0}+\frac{1}{\zeta})}\right] & \text{otherwise}
\end{cases}
\end{equation}
The above are the first-order scattering terms. To compute the higher order terms (fraction of power undergoing $> 1$ scattering events), it is necessary to use the MC method. The slab is discretized into $n=1,\ldots ,\text{N}$ segments of width $\Delta x=L_{x}/\text{N}$. The angular spectrum is also discretized into $m=1,\ldots ,\text{M}$ segments of width $\Delta \theta=2\pi/\text{M}$. Next, the $\text{NM}\times \text{NM}$ ``transition'' matrix $\text{T}(x_n,\theta_m;x_{n^{'}}\theta_{m^{'}})$ is generated, which accounts for the probability of a photon in segment $n$ and directed along $\theta_m$ to scatter in segment $n'$ into $\theta_{m^{'}}$. The $\text{N}\times \text{M}$ ``source'' matrix $\Pi(x_n,\theta_m)$ is defined as the probability distribution of photons in segment $n$ directed along $\theta_{m}$ right after the first scattering event. Lastly, the $\text{NM}\times \text{M}$ ``absorption'' matrix $\text{R}_{T/R}(x_n,\theta_m;\theta_{m^{'}})$ is the probability of a photon to escape the slab via transmission/reflection following its final scattering event in segment $n$ from $\theta_m$ to $\theta_{m^{'}}$. The form for these matrices are as follows. The ``transition'' matrix can be broken into four components:
\begin{equation}
\text{T}(x_n,\theta_m;x_{n^{'}}\theta_{m^{'}}) = p_{\text{esc}}(x_n,\theta_m)\,p_{\text{trvl}}(x_n, \theta_m, x_{n'})\, p_{\text{sct}}(x_{n'},\theta_m)\,\hat{\sigma}(\theta_{m'}-\theta_{m})
\end{equation}
where
\begin{subequations}
\begin{align}
p_{\text{esc}}(x_n,\theta_m) & = \frac{\zeta_m}{\Sigma_{\text{eff}}\Delta x}
			\left(1-e^{-\frac{\Delta x}{\zeta_m}\Sigma_{\text{eff}}}\right)\\
p_{\text{trvl}}(x_n,\theta_m,x_{n'}) & = \begin{cases}
e^{-\frac{x_{n'}-x_n}{\zeta_m}\Sigma_{\text{eff}}} & \quad \frac{x_{n'}-x_n}{\cos{\theta_m}} \geq 0\\
0 & \quad \frac{x_{n'}-x_n}{\cos{\theta_m}} < 0\\
\end{cases}
\\
p_{\text{sct}}(x_{n'},\theta_m) & = 1 - e^{-\frac{\Delta x}{\zeta_m}\Sigma_{\text{eff}}}
\end{align}

\end{subequations}
The \emph{escape} probability, $p_{\text{esc}}(x_n,\theta_m)$, is the probability for a photon to travel through segment $n$ without scattering. The \emph{travel} probability, $p_{\text{trvl}}(x_n,\theta_m,x_{n'})$, is the probability of traveling between segments $n$ and $n'$ without scattering. Note that $p_{\text{trvl}}$ is set to zero in cases where the photon in segment $n$ with $\theta_m$ is oriented such that it is traveling away from $n'$. The \emph{scatter} probability, $p_{\text{sct}}(x_{n'},\theta_m)$, is the probability of scattering within segment $n'$. Lastly, $\hat{\sigma}(\theta_{m'}-\theta_{m})$ is the probability of the photon rotating from $\theta_m$ to $\theta_{m'}$ given that it undergoes a scattering event. The ``source'' matrix is
\begin{equation}
\Pi(x_n,\theta_m) = \hat{\sigma}(\theta_m-\theta_{0})C^{-1} \times \begin{cases}
\frac{\zeta_m}{\zeta_m-\zeta_0}
e^{-(\frac{x_n}{\zeta_0}+\frac{\Delta x}{\zeta_m})\Sigma_{\text{eff}}}
\left(1-e^{-\Sigma_{\text{eff}}\Delta x (\frac{1}{\zeta_{0}}-\frac{1}{\zeta_m})}\right)& \cos{\theta_m} > 0\\
\frac{\zeta_0 \zeta_m}{\zeta_0 + \zeta_m}e^{-\frac{x_n}{\zeta_0}\Sigma_{\text{eff}}}\left(1 - e^{-\Sigma_{\text{eff}}\Delta x (\frac{1}{\zeta_0}+\frac{1}{\zeta_m})}\right) & \cos{\theta_m} < 0
\\
\frac{\Sigma_{\text{eff}} \Delta x }{\zeta_0} e^{-\frac{x_n + \Delta x}{\zeta_0}\Sigma_{\text{eff}}} & \theta_m = \theta_0
\end{cases}
\end{equation}
\begin{equation}
C = \frac{\zeta_m}{\Sigma_{\text{eff}}\Delta x} \left(1-e^{-\frac{\Delta x}{\zeta_m}\Sigma_{\text{eff}}}\right)
\end{equation}
where the coefficient $C$ is required to properly volume-average the source over segment $n$. The transmission and reflection ``absorption'' matrices are
\begin{equation}
\text{R}_{T}(x_n,\theta_m;\theta_{m'}) = \hat{\sigma}(\theta_{m'}-\theta_{m}) \times \begin{cases}
0 & \cos{\theta_{m'}} < 0
\\
\frac{\zeta_{m'}}{\zeta_{m'}-\zeta_m}e^{-\Sigma_{\text{eff}}(\frac{L_x}{\zeta_{m'}} - \frac{x_n}{\zeta_{m}})}\times \\ \left(e^{-\Sigma_{\text{eff}}x_n (\frac{1}{\zeta_{m'}} - \frac{1}{\zeta_{m}})} - e^{-\Sigma_{\text{eff}}L_x (\frac{L_x}{\zeta_{m'}} - \frac{x_n}{\zeta_{m}})}\right) & \cos{\theta_{m}} > 0
\\
\frac{\zeta_{m'}}{\zeta_m+\zeta_{m'}}e^{-\Sigma_{\text{eff}}(\frac{L_x}{\zeta_{m'}}+\frac{x_n}{\zeta_m})}\times \\ \left(e^{\Sigma_{\text{eff}}x_n(\frac{1}{\zeta_m}+\frac{1}{\zeta_{m'}})} - e^{\Sigma_{\text{eff}}L_x(\frac{1}{\zeta_m}-\frac{1}{\zeta_{m'}})}\right) & \cos{\theta_{m}} < 0
\\
\frac{\Sigma_{\text{eff}}}{\zeta_m}e^{-\Sigma_{\text{eff}}(\frac{L_x}{\zeta_{m'}}-\frac{x_n}{\zeta_m})}(L_x - x_n) & \theta_m = \theta_{m'}
\end{cases}
\end{equation}

\begin{equation}
\text{R}_{R}(x_n,\theta_m;\theta_{m'}) = \hat{\sigma}(\theta_{m'}-\theta_{m}) \times \begin{cases}
0 & \cos{\theta_{m'}} > 0
\\
\frac{\zeta_{m'}}{\zeta_{m}-\zeta_{m'}}e^{-\Sigma_{\text{eff}}\frac{x_n}{\zeta_{m}}}\left(1 - e^{-\Sigma_{\text{eff}}x_n (\frac{1}{\zeta_{m'}} - \frac{1}{\zeta_{m}})}\right) & \cos{\theta_{m}} < 0
\\
\frac{\zeta_{m'}}{\zeta_m+\zeta_{m'}}e^{\Sigma_{\text{eff}}\frac{x_n}{\zeta_{m'}}}\left(e^{-\Sigma_{\text{eff}}x_n(\frac{1}{\zeta_m}+\frac{1}{\zeta_{m'}})} - 1\right) & \cos{\theta_{m}} > 0
\\
\frac{\Sigma_{\text{eff}}}{\zeta_m}e^{-\frac{\Sigma_{\text{eff}}}{\zeta_{m'}}x_n}x_n & \theta_m = \theta_{m'}
\end{cases}
\end{equation}
 Using these matrices, one can calculate the higher-order transmitted/reflected wave-spectrum terms:
\begin{subequations}
\begin{align}
P_{T/R,2}(\theta_m) & = \Pi \cdot \text{I} \cdot \text{R}_{T/R}\\
P_{T/R,l}(\theta_m) & = \Pi \cdot \text{T}^{l-2} \cdot \text{R}_{T/R}  \quad \text{for } l \geq 3
\end{align}
\end{subequations}
where $l$ is the order of the scattering term, and $\text{I}$ is the $\text{NM}\times \text{NM}$ identity matrix. In summing all scattering terms, the total transmitted and reflected wave-spectrum is 
\begin{subequations}
\begin{align}
P_{T}(\theta_m) - \frac{P_{ball}}{\Delta \theta}\delta_{\theta_m,\theta_0} & = P_{T,1}(\theta_m) + \sum_{l=1}^{\infty} P_{T,l}\\
P_{R}(\theta_m) & = P_{R,1}(\theta_m) +\sum_{l=1}^{\infty} P_{R,l}
\end{align}
\end{subequations}
and $\delta_{i,j}$ is the Kronecker delta. Furthermore, eq. (21a) can be rewritten as:
\begin{equation}
P_{T}(\theta_m) - \frac{P_{ball}}{\Delta \theta}\delta_{\theta_m,\theta_0} - P_{T,1}(\theta_m)= \Pi \cdot \left( \text{I} + \sum_{l=1}^{\infty}\text{T}^l \right) \cdot \text{R}_{T} = \Pi \cdot \left( \text{I} - \text{T}\right)^{-1}\cdot \text{R}_{T}
\end{equation} 
A similar form applies to eq. (21b). The second relation in eq. (22) produces the solution following a matrix inversion. In practice, it is often faster to evaluate the first relation and truncate the series at a finite $l$ when the solution is sufficiently converged \cite{yang2018markov}.

In deriving $p_{\text{esc}}$ and $p_{\text{sct}}$, the possibility of multiple scattering events in segment $n$ is neglected. This is a reasonable assumption as long as $\Delta x \equiv \frac{L_x}{\text{N}} \ll \frac{\zeta}{\Sigma_{\text{eff}}}$. The population of photons with $\zeta \approx 0$ is the largest source of error for any finite N. Nevertheless, in practice, $P_{T/R}$ is found to converge as long as $\Sigma_{\text{eff}} \Delta x \ll 1$. A criteria for the angular resolution is not as straight-forward. It depends on the smoothness of $\hat{\sigma}(\theta)$. Naturally, a fine resolution is needed to accurately resolve sharp peaks in $\hat{\sigma}(\theta)$.

\section{Verification of SAS-MC with numeric full-wave solver}
\label{sec:6}
The SAS-MC model is compared with the higher-fidelity finite-element full-wave code PETRA-M \cite{shiraiwa2017rf}. First, the SAS model for a single filament is compared to PETRA-M. Then the same is done for the MC model, which accounts for multiple filaments in a slab.

\subsection{Scattered field for single filament}
\label{sec:6.1}
\begin{figure}[!h]
\centering
\includegraphics[width=16cm, height=8cm]{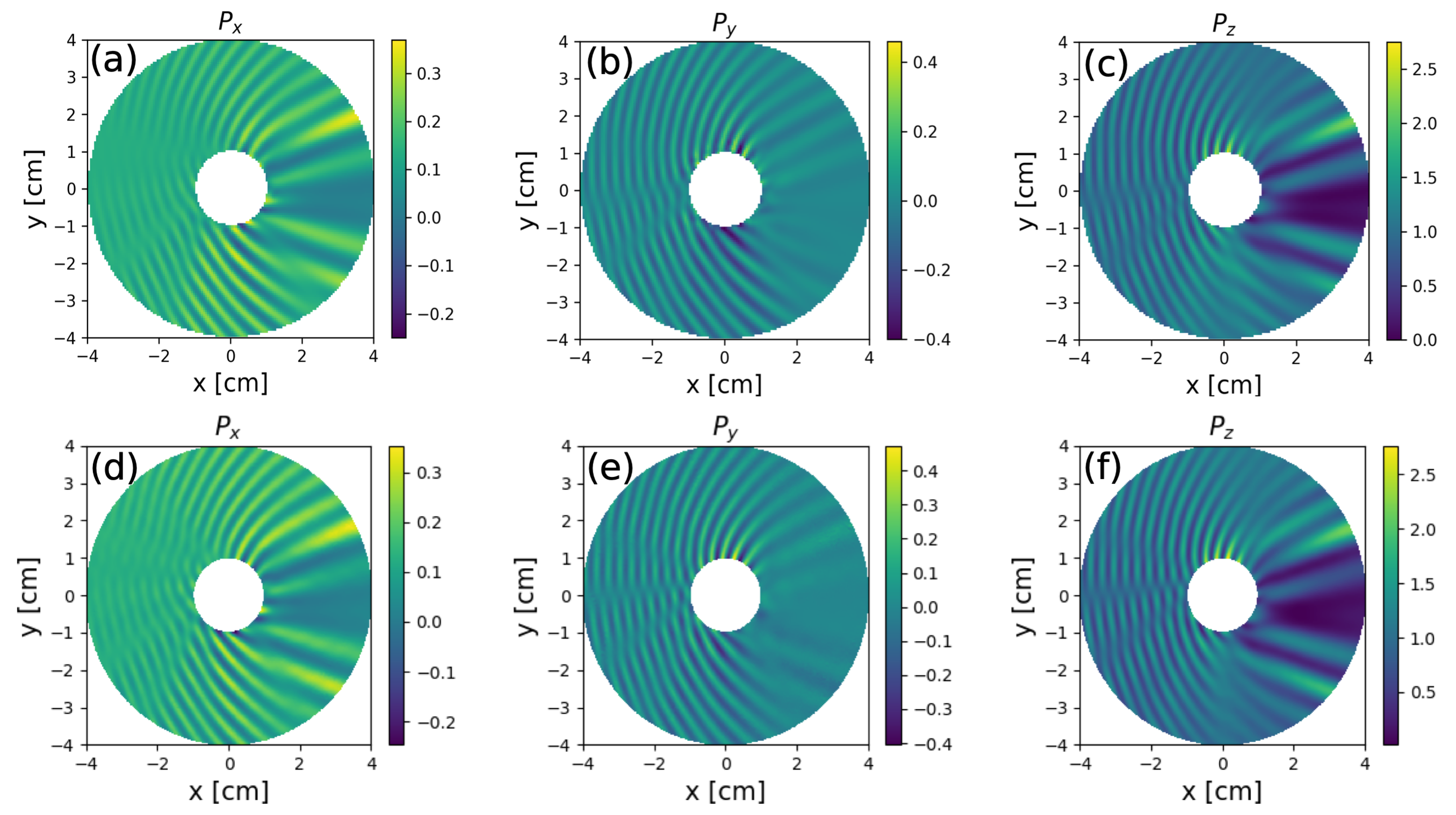}
\caption[font=5]{Gaussian filament: Normalized Poynting flux computed using the SAS model (a-c) and PETRA-M (d-f). $n_0=1\times 10^{19} \, \text{m}^{-3}$, $\text{B}=4 \, \text{T}$, $f=4.6\, \text{GHz}$, $N_{||} =2$, $n_b/n_0 = 4.8$, and $a_b=1\, \text{cm}$.}
\label{fig:2}
\end{figure}

Consider an incident slow wave with a prescribed frequency $f$ and parallel refractive index $N_{||}\equiv\frac{c k_{||}}{\omega}$. Also assume a filament with Gaussian radial profile such that
\begin{equation}
n(\rho) -n_0= n_0 \left( \frac{n_b}{n_0} -1 \right) e^{-\left(\frac{2\sqrt{\text{ln}(2)}\rho}{a_b}\right)^{2}}
\end{equation}
where $n_{b}/n_0$ is the relative density at the filament's peak ($\rho = 0$), and $a_b$ is re-defined as the \emph{full-width half-max} of the filament. A case with $n_0=1\times 10^{19}\, \text{m}^{-3}$, $\text{B}=4\, \text{T}$, $f=4.6\, \text{GHz}$, $N_{||} =2$, $n_b/n_0 = 4.8$, and $a_b=1\, \text{cm}$ is simulated using the SAS model. Simulation resolution is $R=22$ and $M=100$. Figure \ref{fig:2}(a-c) show the (x,y,z) components of the time-averaged Poynting flux $\bold{S}$ exterior to the filament. $\bold{S}$ is calculated using the relation $\bold{S} = \frac{1}{2 \mu_0 \omega}\Im{\bold{E}^{*} \cross (\nabla \cross \bold{E})}$. The normalized field $\bold{P} \equiv \frac{\bold{S}}{|\bold{S}_{inc}|}$ is introduced, where $\bold{S}_{inc}$ is the Poynting flux of the incident wave. Figure \ref{fig:2} reveals a shadowing effect downstream of the filament. The striations in the field indicate strong back and side-scattering of the incident wave. Note that $\bold{P}$ has been numerically computed from the interpolated $\bold{E}$-field on a grid in the (x,y)-plane. This method of plotting $\bold{P}$ is susceptible to large errors inside the filament where the gradients of $\bold{E}$ are large. Therefore, $\bold{P}$ inside the filament is not plotted.

Figures \ref{fig:2}(d-f) show this case repeated in PETRA-M, and result in excellent agreement with the SAS model. This 2D simulation is done in a circular domain. The incident wave is excited to the left of the filament using an external current source term. To approximate an infinite background plasma, a perfectly-matched layer (PML) is modeled at the perimeter of the circular simulation domain.
 
\subsection{Reflection coefficient for turbulent slab}
\label{sec:6.2}
Next, the Markov chain (MC) step of the SAS-MC model is compared with a turbulent slab modeled in PETRA-M. This will reveal whether the far-field limit, a critical approximation in the MC model, is valid for the treatment of LH scattering in the SOL. In theory, the far-field approximation should break down as filaments are packed closer together \cite{mishchenko2014electromagnetic}.

\begin{figure}[!h]
\centering
\includegraphics[scale=0.8]{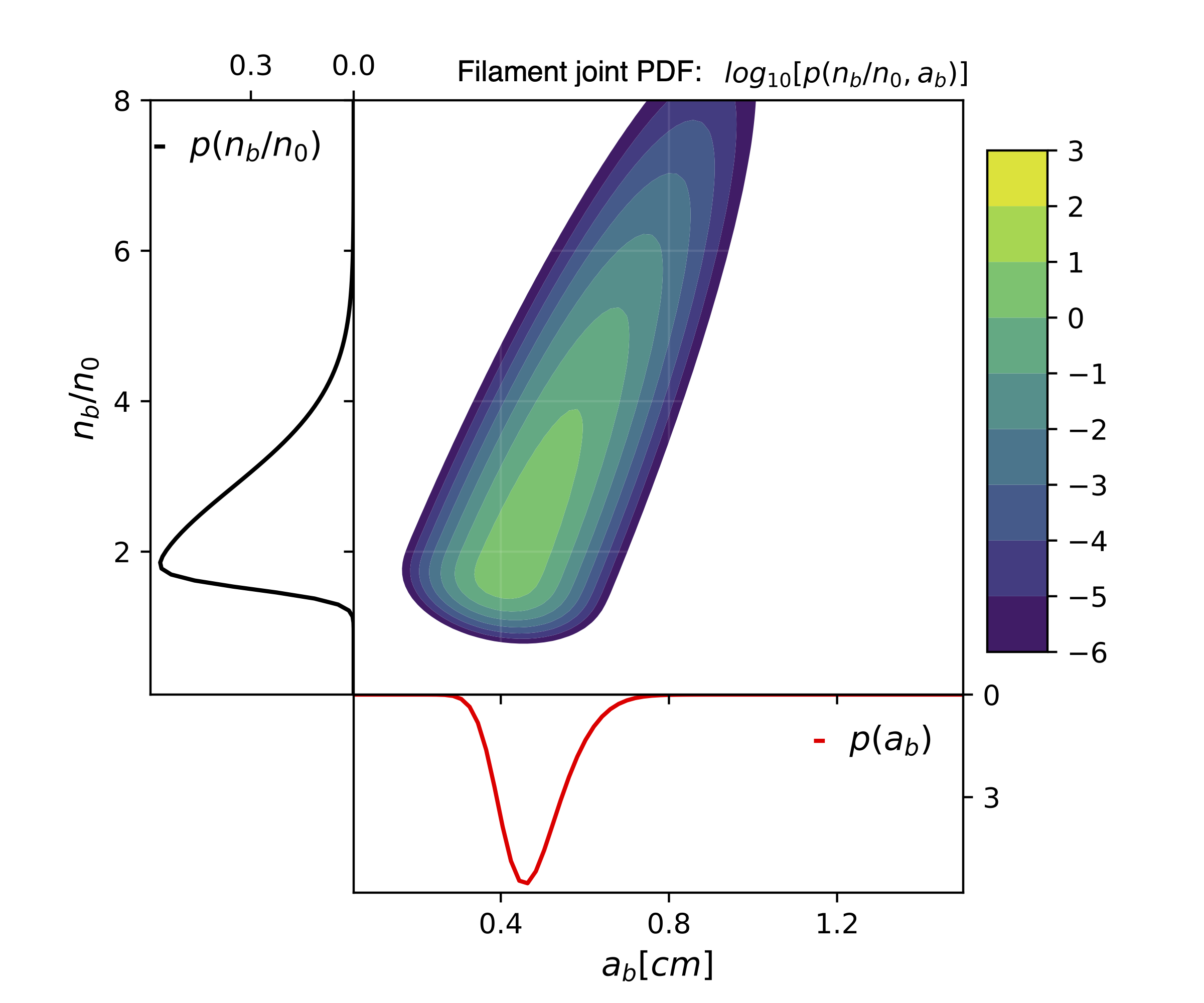}
\caption[font=5]{Example joint-PDF of filament parameters $a_b$ and $n_b/n_0$. A skewed-normal distribution is assumed for $a_b$ and $n_b/n_0$. $\langle a_b \rangle = 0.48\,$cm and $\langle n_b/n_0 \rangle = 2.6$. Filament size and density shape parameters are +10 and +7, respectively. Filament size and density standard deviation are 0.1$\,$cm and 1.35, respectively. Bi-variate correlation coefficient is 0.9.}
\label{fig:7}
\end{figure}

For the MC model, $\sigma_{\text{eff}}(\theta)$ must be calculated, which first requires prescribing a joint-PDF of filaments. Figure \ref{fig:7} plots an example joint-PDF. A skewed-normal distribution is assumed for $a_b$ and $n_b/n_0$. In this case, $\langle a_b \rangle = 0.48\,$cm and $\langle n_b/n_0 \rangle = 2.6$. These values are bounded by experimental SOL measurements \cite{terry2003observations, zweben2002edge}. (Assuming SOL turbulence is predominantly filamentary, the approximation $\langle n_b/n_0 \rangle \approx 1 + \frac{n_{\text{RMS}}}{n}f_{p}^{-1}$ is made, where $f_{p}$ is the packing fraction and is assumed to be $0.2$). Filament size and density skewness are prescribed via the shape parameter for a skewed normal distribution. These are chosen to be +10 and +7, respectively. Filament size and relative density standard deviation are 0.1$\,$cm and 1.35, respectively. The bi-variate correlation coefficient is 0.9.

\begin{figure}[!h]
\centering
\includegraphics[scale=0.6]{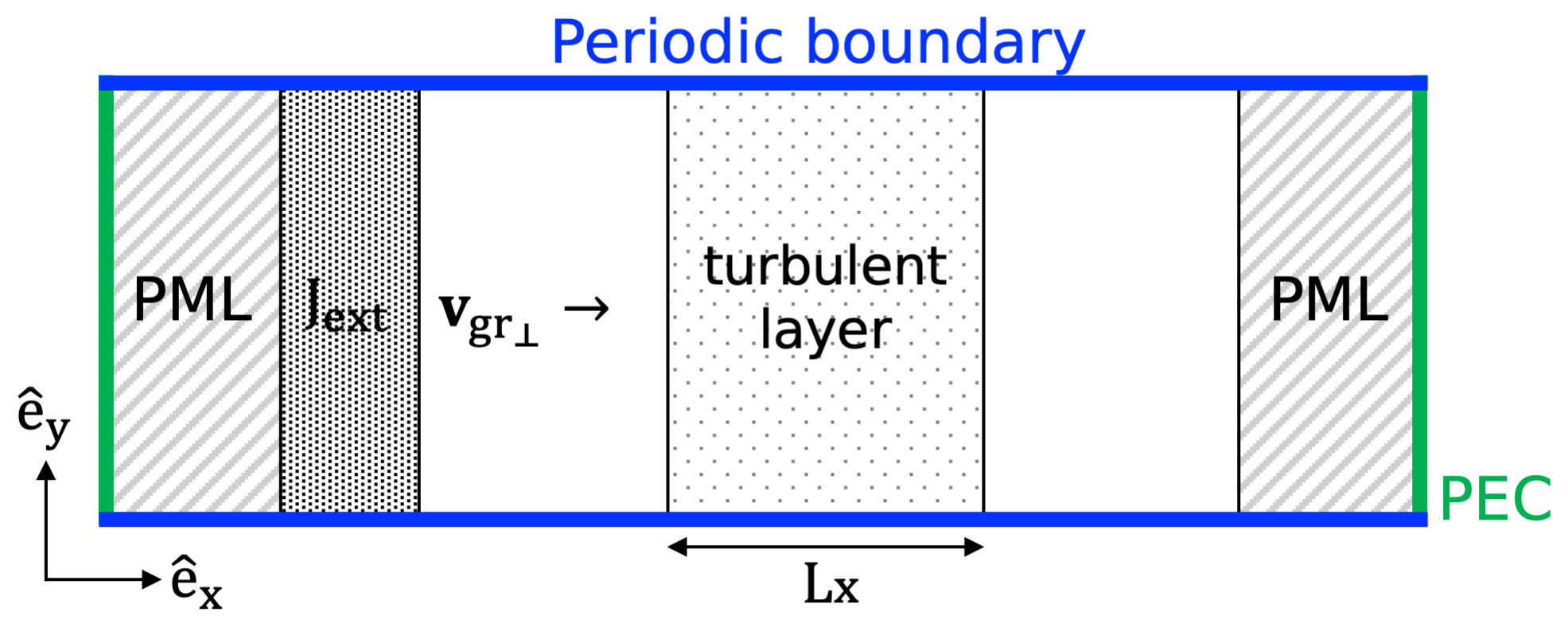}
\caption[font=5]{Setup for PETRA-M simulation with turbulent slab. ``PML'' denotes perfectly matching layer. ``PEC'' denotes perfect electric conductor.}
\label{fig:11}
\end{figure}

Figure \ref{fig:11} shows the simulation setup for a slab turbulent geometry in PETRA-M. A slow wave traveling in the $+x$-direction interacts with a turbulent layer populated with Gaussian filaments. The filaments are randomly generated in the slab using a Monte-Carlo approach\cite{sierchio2016comparison,biswas2020study}. Each filament is generated with a randomly picked $n_b/n_0$ and $a_b$ with a probability that satisfies the prescribed joint-PDF $p(n_b/n_0, a_b)$. In this way, the turbulent slab used in the SAS-MC model and in PETRA-M are made statistically equivalent. The incident slow wave is excited with an external current density source function upstream of the turbulence. Top and bottom boundaries are periodic. The turbulence is also periodic in the $y$-direction. In order to minimize the periodic geometry's effect on the wave, the $y$ length of the solution domain is much larger than $\langle a_b \rangle$ or $k_{inc\perp}$. To mimic infinite domain in the $\pm x$-direction, the left and right boundaries can be modeled as either a perfectly matched layer (PML) or an absorbing boundary condition (ABC). The PML, while more computationally efficient, does not work when both the slow and fast wave can propagate in the background plasma.

In PETRA-M, the reflection coefficient $F_{\text{ref}}=1-P_{x}/P_{x,0}$ is calculated and directly compared with the SAS-MC value. $P_{x}$ is the x-component of the Poynting flux and $P_{x,0}$ is the ``nominal'' value when no turbulence is present. In the SAS-MC model, $F_{\text{ref}}=\int P_{\text{R}}(\theta)d\theta$.

Table \ref{table:1} compares $F_{\text{ref}}$ computed in PETRA-M and the SAS-MC model. Both models follow the same general trend. Cases (1-4) and (8-11) reveal that $F_{\text{ref}}$ increases with $f_{p}$. Cases (3,5,6,8) reveal that $F_{\text{ref}}$ increases with $\langle n_b/n_0 \rangle$ and decreases with $\langle a_b \rangle$. This is consistent with previous scattering theories\cite{andrews1983scattering,ott1979lower}. Another way to analyze the trends in $F_{\text{ref}}$ between models is by inspecting $\Sigma_{\text{eff}}L_{x}$ calculated in the SAS-MC model. This is the attenuation factor for the ballistic power (see eq. (11)). As $\Sigma_{\text{eff}}L_{x}$ increases, so should $F_{\text{ref}}$. Indeed, this is true for both models.

In general, the SAS-MC model over-predicts $F_{\text{ref}}$, such that the absolute error $\Delta F_{\text{ref}} \equiv F_{\text{ref,SAS-MC}} - F_{\text{ref,PETRA-M}} \geq 0$. This error increases with $f_p$. Again, this is due to the far-field approximation breaking down. While far-field validity is dependent on $f_{p}$, the aggregate error depends on $\Sigma_{eff} L_{x}$. For example, the SOL width is varied between cases (2) and (7), while the turbulence is kept statistically identical. The $L_{x}=5\,$cm case results in $\Delta F_{\text{ref}} = 0.00$. For the $L_{x}=15\,$cm case, $\Delta F_{\text{ref}} = +0.13$.

SOL measurements indicate $f_p \approx 0.05-0.25$ \cite{agostini2007study,carralero2018role,zweben2011estimate}. SOL widths are also $<5$ cm in present-day devices. As a result, cases (2) and (9) are most representative of a C-Mod SOL, depending on whether the background density is evaluated at the far-SOL or the separatrix, respectively. At low-density (case 2), the two models agree well ($\Delta F_{\text{ref}} = 0.00$). At high density (case 9), the SAS-MC model over-predicts $F_{\text{ref}}$, such that $\Delta F_{\text{ref}} = +0.08$. A possible reason for the disagreement at high density may be because $\Sigma_{\text{eff}}L_{x}$ is greater (compared to similar cases at low density).

\begin{table}[!h]
\begin{tabular}{c|ccccc|c|cc}
\multicolumn{1}{c|}{} &
\multicolumn{5}{c|}{Plasma parameters} & 
\multicolumn{1}{c|}{} & 
\multicolumn{2}{c}{$F_{\text{ref}}$} \\
\hline
Case \# &
$n_{0}\times10^{19}\,[\text{m}^{-3}]$ & 
$\langle n_{b}/n_{0} \rangle$ & 
$\langle a_b \rangle\,$[cm]    & 
$f_{p}$       & 
$L_{x}\,$[cm]      & 
$\Sigma_{\text{eff}}L_{x}$ &
Petra-M       & 
SAS-MC       \\ 
\hline
1			&0.55  					& 2.60            		& 0.48          	 	& 0.02 						&5.0						& 0.26		& 0.01          	& 0.02         \\
2			&0.55  					& 2.60                 & 0.48          	 	& \textbf{0.10} 		&5.0						& 1.29		& 0.13          	& 0.13         \\
3			&0.55  					& 2.60                 & 0.48          	 	& \textbf{0.25} 		&5.0						& 3.23		& 0.18          	& 0.31         \\
4			&0.55  					& 2.60                 & 0.48          	 	& \textbf{0.50} 		&5.0						& 6.45		& 0.29          	& 0.48         \\
5			&0.55  					& 2.60                 & \textbf{1.10}		& 0.25          	  		&5.0						& 1.33		& 0.02          	& 0.06         \\
6			&0.55                	& \textbf{1.80}    	& 0.48          	 	& 0.25          	  		&5.0						& 1.42		& 0.04          	& 0.13         \\
7			&0.55  					& 2.60                 & 0.48          	 	& 0.10						&\textbf{15.0}		& 3.89		& 0.22          	& 0.35         \\
8			&\textbf{2.25}   	& 2.60                 	& 0.48          	 	& 0.02          	  		&5.0						& 0.40		& 0.04             & 0.04         \\        
9			&2.25				   & 2.60                 	& 0.48          	 	& \textbf{0.10}        &5.0						& 2.02		& 0.15             & 0.23         \\
10		&2.25				   & 2.60                 	& 0.48          	 	& \textbf{0.25}        &5.0						& 5.06		& 0.34             & 0.46         \\
11		&2.25				   & 2.60                 	& 0.48          	 	& \textbf{0.50}        &5.0						& 10.1		& 0.55             & 0.65     
\end{tabular}
\caption[font=5]{Comparison between SAS-MC model and PETRA-M. Slow-wave launched at $4.6\, \text{GHz}$ and $N_{||}=2$ with $\text{B}=4\, \text{T}$. Filament joint-PDF parameters are same as in Figure \ref{fig:7} unless otherwise noted. $\Sigma_{\text{eff}}L_{x}$ calculated in SAS-MC model. For each case, results from multiple iterations (with different turbulence realizations) are averaged until $F_{\text{ref}}$ is statistically converged.}
\label{table:1}
\end{table}

\subsection{Comments on computational cost}
\label{sec:6.3}
Generally, the semi-analytic scattering method has three key advantages compared to finite-element Maxwell solvers.  (1) The large (in fact infinite) background plasma region does not need to be meshed. (2) It exactly solves scattering problems, since the infinite exterior domain does not need to be artificially truncated. (3) Analyzing the scattered wave-spectrum is straight-forward, since the solution is already deconvolved into the constituent poloidal mode-numbers \emph{for each} branch.

Points 1 and 2 result in the SAS-MC model being considerably less expensive than slab turbulence simulations in PETRA-M. Using the SAS technique, computing a single differential scattering-width $\sigma(\theta;n_b/n_0, a_b )$ takes $\sim\! 10$ seconds on a single CPU, and considerably less time if parallelized between poloidal mode-numbers. Computing $\sigma_{\text{eff}}(\theta)$ may require sampling a few hundred combinations of $(n_b/n_0, a_b)$, depending on the filament joint-PDF. Fortunately, each sampled $\sigma(\theta;n_b/n_0, a_b )$ needs to be computed only once. Any number of $\sigma_{\text{eff}}(\theta)$ can then be generated from the sampled differential scattering-widths.

The most expensive process in the MC routine is generating the transition matrix T. This takes $\sim 20$ seconds on a single CPU, depending on poloidal and radial bin resolution.

Using PETRA-M, each $n_0 = 2.25 \times 10^{19}\, \text{m}^{-3}$, $L_{x} = 5\,$cm slab case required $\sim\!25\,$CPU-hours and $\sim\! 300\,$GB of RAM on the MIT Engaging computing cluster. The size of PETRA-M simulations is primarily limited by available RAM. In comparison, all computations for the SAS-MC model have been conducted on a PC with $8\,$GB of available RAM.

\subsection{Caveats to the SAS-MC model}
\label{sec:6.4}
The SAS-MC model offers higher physics-fidelity than ray-tracing, while being computationally less expensive than numeric full-wave solvers. This is possible due to a number of assumptions made in the model that makes it less universally applicable than numeric full-wave solvers.

The semi-analytic scattering (SAS) model assumes a homogeneous background plasma with a cylindrical scattering object (the filament) that is poloidally and azimuthally symmetric. This allows an efficient solution scheme using separation of variables. In reality, filaments usually develop a shock front as they convect outward \cite{krasheninnikov2008recent}, and the resulting crescent-like filament shape can lead to significantly modified scattering behavior, at least for ion-cylcotron waves \cite{tierens2020importance}. Furthermore, the filament in the SAS model is assumed to be aligned with the magnetic field, so that $\nabla_{||} \left( \frac{n}{n_0} \right)=0$. This is likely a reasonable assumption since filaments introduce a $\nabla_{||} \left( \frac{n}{n_0} \right)$ that is much smaller than $k_{||}$ of the LH wave\cite{grulke2014experimental}. As a result, the effect of $k_{||}$ broadening due to a typical SOL filament is small \cite{madi2015propagation}.

The Markov chain (MC) model introduces additional assumptions. The SOL is treated as a slab, which means the effect of toroidal geometry is neglected. This is a reasonable assumption for the treatment of first-pass scattering in front of the antenna. In addition, the background plasma and turbulence parameters are constant within the slab, when in reality they are sensitive to the radial coordinate in a tokamak. The MC model also assumes the reflected wave-spectrum is lost, when in reality a fraction of this power may once again reflect at a cutoff and re-enter the core plasma. Lastly, the MC model assumes the filaments are far enough apart so that the RTE is valid. This assumption is increasingly poor as $f_p$ rises.

In order to quantify the inaccuracies introduced by the MC model, it may be worth-while to model scattering along the full extent of a ray-trajectory in a realistic tokamak geometry. This can be done by employing $\sigma(\theta; n_0, B, N_{||}, \langle n_b/n_0 \rangle, \langle a_b \rangle)$ as a scattering probability in a Monte-Carlo ray-tracing simulation, similar to what has been done for the $k$-scattering model \cite{bonoli1982toroidal,bertelli2013effects}. Alternatively, 3D PETRA-M simulations of LH launch in a turbulent SOL, with realistic geometry, would address all the caveats mentioned. These tests are outside the scope of this paper.

\section{SAS-MC applied to Lower Hybrid scattering}
\label{sec:7}
\begin{figure}[!h]
\centering
\includegraphics[scale=0.6]{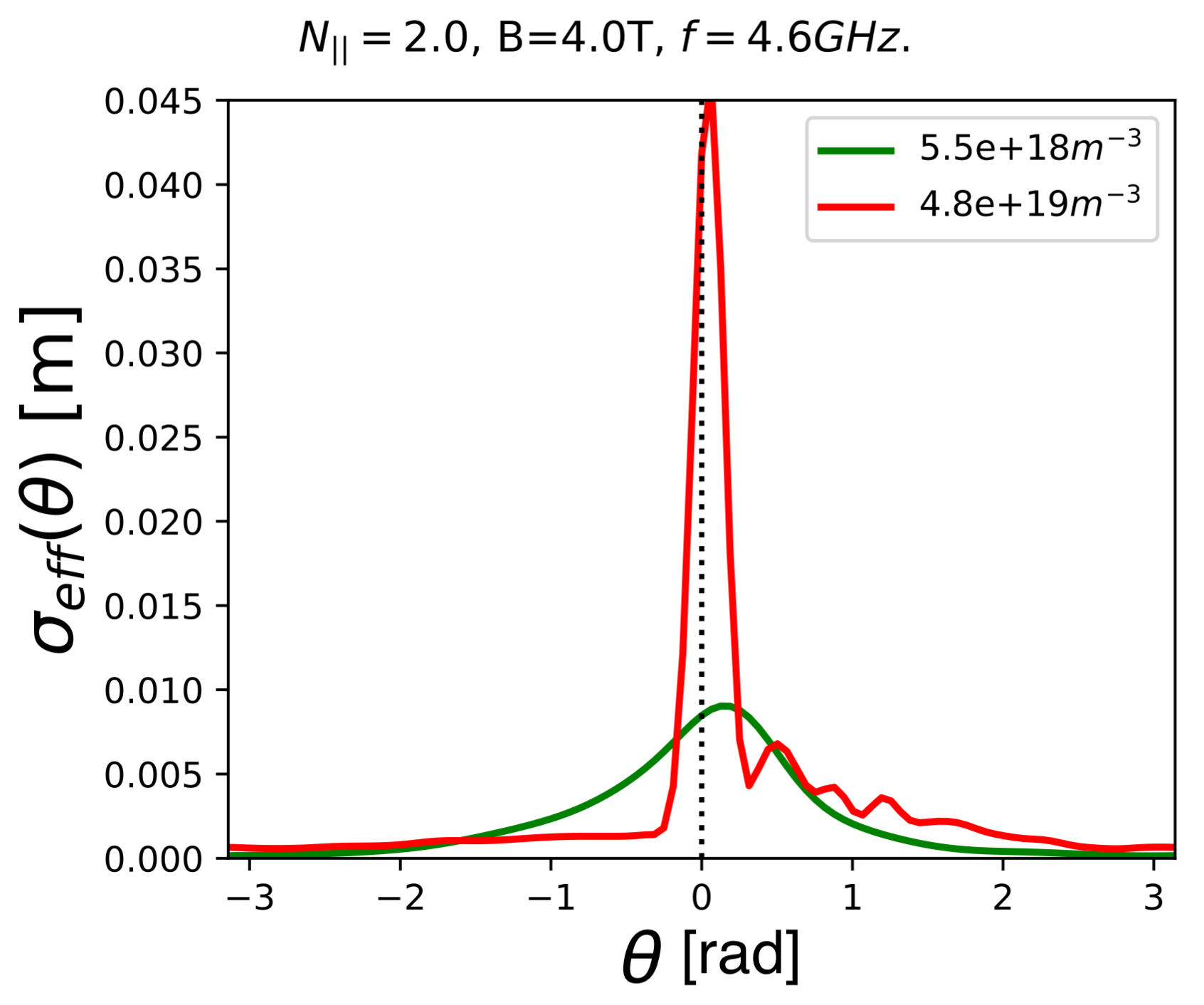}
\caption[font=5]{Effective differential scattering-width $\sigma_{\text{eff}} (\theta)$. Calculated using filament joint-PDF shown in Figure \ref{fig:7}. Green and red lines denote low and high background density, respectively.}
\label{fig:8}
\end{figure}

The SAS-MC model is applied to LH wave scattering in front of the antenna in C-Mod. Figure \ref{fig:8} plots $\sigma_{\text{eff}} (\theta)$ for the joint-PDF in Figure \ref{fig:7}, and $N_{||}=2$, $f=4.6\, \text{GHz}$, and $\text{B}=4\, \text{T}$. At low background density ($n_0 = 5.5 \times 10^{18} \text{m}^{-3}$), $\sigma_{\text{eff}} (\theta)$ resembles a wrapped-Cauchy distribution, though slightly skewed so that it peaks at $+0.2\, \text{rad}$. At high background density ($n_0 = 4.8 \times 10^{19} \text{m}^{-3}$), $\sigma_{\text{eff}} (\theta)$ is sharply peaked near $+0.05\, \text{rad}$ and is very asymmetric. A fat-tail exists only for $\theta > 0$. This tail also has fine structures that do not exist in the low density case. $\sigma_{\text{eff}} (\theta)$ is more asymmetric at higher densities because $|\epsilon_{xy}|$ is larger, and therefore the effect of asymmetric scattering for any given filament is stronger (see \hyperref[sec:7.2]{Section 7.2} for more detail).

\subsection{Parametric scan of SOL density and filament parameters}
\label{sec:7.1}
Figures \ref{fig:3}(a,c,e) plot $\sigma_{S}$ as a function of background density $n_0$, relative filament density $n_b/n_0$, and filament width $a_b$. The incident wave is at 4.6$\,$GHz with $N_{||}=2$, typical for LH launch in C-Mod. In accordance with the low-field-side SOL in C-Mod, $\text{B}=4\, \text{T}$. As expected, $\sigma_{S}$ increases as $n_b/n_0$ deviates from unity. Notably, scattering resonances can be seen, as indicated by bands of higher $\sigma_{S}$. These are due to standing wave resonances excited within the filament, which result in stronger coupling to the scattered waves. Expressions for these resonances can be analytically derived for the ``flat-top'' filament case\cite{myra2010scattering, tierens2020filament}, and are related to radial and poloidal harmonics in cylindrical geometry. To the authors' knowledge, these analytic calculations are intractable for Gaussian (and more general) filament cross-sections.

\subsection{Asymmetric scattering}
\label{sec:7.2}
\begin{figure}[!h]
\centering
\includegraphics[scale=0.8]{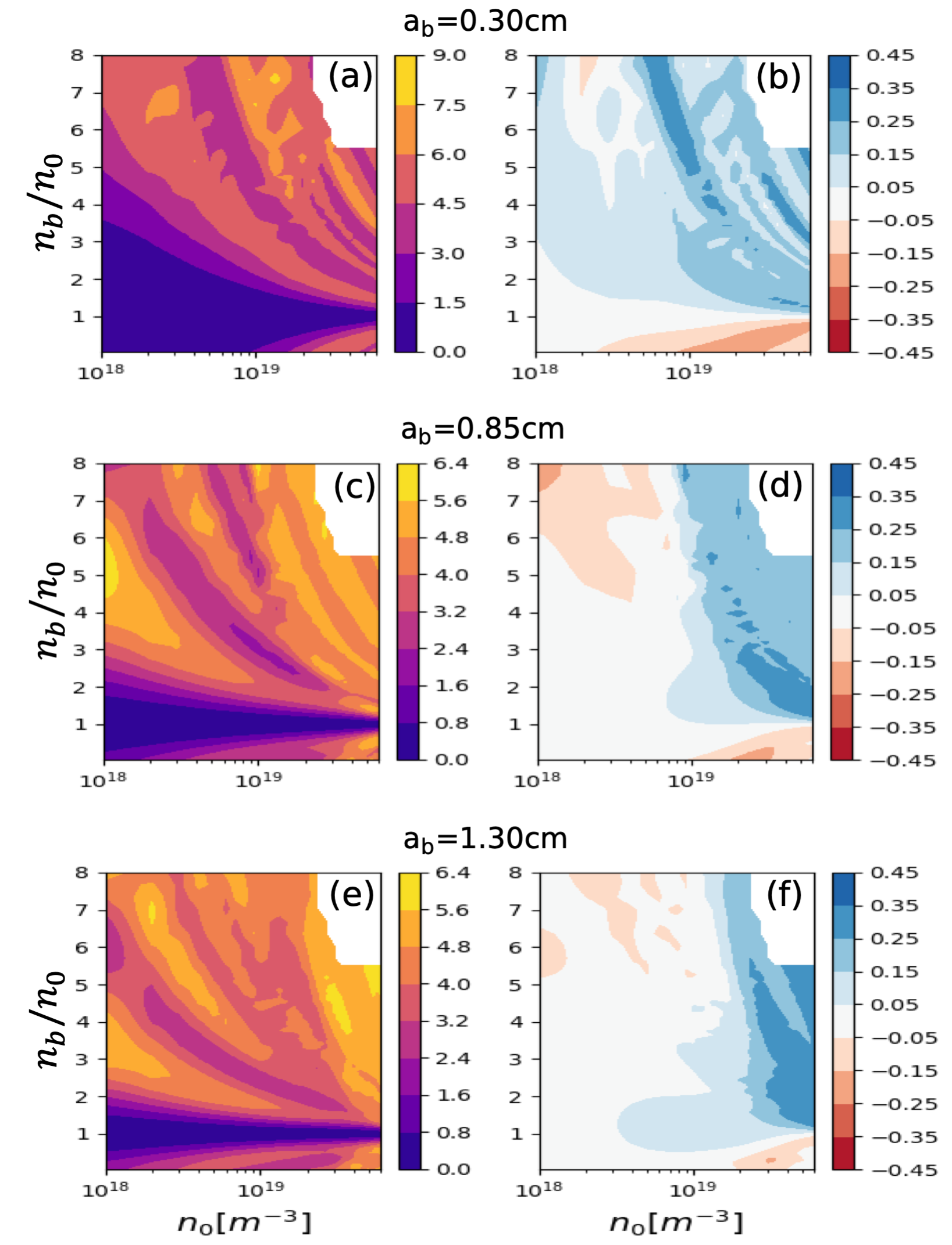}
\caption[font=5]{Plots of (a) scattering-width $\sigma_{S}$ and (b) asymmetric scattering metric $\alpha$ for an incident slow-wave at 4.6$\,$GHz, $N_{||}=2$. $\text{B}=4\, \text{T}$. $a_b =0.30\,\text{cm}$. (c) and (d) plot $\sigma_{S}$ and $\alpha$, respectively, for $a_b=0.85\,\text{cm}$. (e) and (f) plot $\sigma_{S}$ and $\alpha$, respectively, for $a_b=1.30\,\text{cm}$. The white region in the upper-right corner of each subplot has no data plotted.}
\label{fig:3}
\end{figure}

The quantity $\alpha$ is introduced as a metric for asymmetric scattering.
\begin{equation}
\alpha = \int_{0}^{\pi} \hat{\sigma}_{S}(\theta) d \theta - \frac{1}{2}
\end{equation}
$\alpha=-0.5,+0.5$ denote that power is only scattered downwards ($-\pi < \theta < 0$), upwards ($0 < \theta < \pi$). If $\alpha=0$ then an equal fraction of power is scattered downwards and upwards. Figures \ref{fig:3}(b,d,f) reveal that $\alpha$ is not guaranteed to be zero. This means, in general, $\sigma (\theta)$ is not an even function, and the scattered power is not equally distributed downwards or upwards. This effect is most noticeable at high background densities ($n_0 \gtrsim 2\times 10^{19} \text{m}^{-3}$). For positive density modifications ($n_b/n_0 > 1$), power is predominantly scattered upwards. The reverse is true for negative density modifications ($n_b/n_0 < 1$). In the typical tokamak SOL, filaments are predominantly denser than the background plasma\cite{graves2005self}. As a result, the incident wave is preferentially scattered upwards. The strength of this asymmetry depends on the statistical properties of the filaments.

Asymmetric scattering is possible in an anisotropic medium. Specifically, the dielectric dyadic tensor, $\epsilon$, has off-diagonal components $\epsilon_{12} =  -\epsilon_{21} \neq 0$, which permits asymmetric scattering\cite{wu1994scattering}. $\epsilon_{12}=-i \epsilon_{xy}$. In the LH limit, $\epsilon_{xy}\approx \omega_{pe}^2/{\omega \Omega_{ce}}$, where $\Omega_{ce} \equiv eB/m_e$ is the electron cyclotron frequency. It is clear that the sign of $\epsilon_{xy}$ is dependent on the sign of $B$ (that is, whether the magnetic field is oriented co-parallel or counter-parallel with $\hat{e}_{z}$). Correspondingly, when the direction of $\bold{B}$ is flipped, $\epsilon \rightarrow \epsilon^{T}$, and $\sigma (\theta) \rightarrow \sigma (-\theta)$ .

Notably, this asymmetric scattering effect is not accounted for in previous treatments for LH wave scatter. It is easy to see why this is the case for models that assume drift-wave like turbulence. These models assume that density fluctuations are equally likely to be above or below the background density. Therefore, this asymmetric scatter effect is statistically canceled out.

There also exist LH scattering models that assume coherent turbulent structures that can, on average, be denser than the background \cite{hizanidis2010fokker,biswas2020study}. These models also do not account for asymmetric scattering, because they make the ray-tracing approximation. 

The reason why ray-tracing cannot model asymmetric scattering is subtle. It is related to the breakdown of the ray-tracing approximation. Consider the ray-tracing equations for a LH ray initially propagating with $\bold{N}_{\perp}$ aligned along the x-direction, and background $\bold{B}$ aligned along the z-direction. The ray-tracing equations involve partial derivatives of $\text{det}(\bold{D})$, where $\bold{D}$ is now the dielectric tensor and 
\begin{equation}
\bold{D} \cdot \bold{\tilde{E}}=
\begin{bmatrix}
\epsilon_{\perp} - N_{||}^2 &  -i\epsilon_{xy} &  N_{\perp}N_{||} \\

i\epsilon_{xy} & \epsilon_{\perp} - N^2 & 0 \\

N_{\perp}N_{||} & 0 & \epsilon_{||} - N_{\perp}^2 

\end{bmatrix}
\cdot 
\begin{bmatrix}
\tilde{E_x} \\ 
\tilde{E_y} \\
\tilde{E_z} 
\end{bmatrix}
= 0
\end{equation}
where $\bold{\tilde{E}}(\bold{r})$ is the slowly-varying part of the electric field. det($\bold{D}$) has terms that are quadratic in $\epsilon_{xy}$, but no linear $\epsilon_{xy}$ terms. As a result, information about the sign of $B$ along $\hat{e}_{z}$ is lost. Compare this to the full EM wave-equation, with no ray-tracing approximation, written in a form similar to that of eq. (25).
\begin{equation}
\begin{bmatrix}
\epsilon_{\perp} - \frac{c^2}{\omega^2}(F_{yy} + F_{zz}) &  \frac{c^2}{\omega^2}F_{yx} - i\epsilon_{xy} &  \frac{c^2}{\omega^2}F_{zx} \\

\frac{c^2}{\omega^2}F_{xy} + i\epsilon_{xy} & \epsilon_{\perp} - \frac{c^2}{\omega^2}(F_{zz}  + F_{xx}) & \frac{c^2}{\omega^2}F_{zy} \\

\frac{c^2}{\omega^2}F_{xz} & \frac{c^2}{\omega^2}F_{yz} & \epsilon_{||} - \frac{c^2}{\omega^2}(F_{xx} + F_{yy}) 

\end{bmatrix}
 \cdot 
\begin{bmatrix}
\tilde{E_x} \\ 
\tilde{E_y} \\
\tilde{E_z} 
\end{bmatrix}
= 0
\end{equation}
where
\begin{subequations}
\begin{align}
F_j = k_j - i \pdv{}{j} \\
F_{jl} = F_j(F_l)
\end{align}
\end{subequations}
Equation (26) accounts for $\nabla \bold{k}$ and $\nabla \bold{\tilde{E}}$ terms which are usually neglected in ray-tracing because the plasma is assumed sufficiently homogeneous, such that $k_{\perp}L_n \gg 1$, where $L_n\equiv |\frac{\nabla n}{n_0}|^{-1}$ is the characteristic length of the density in-homogeneity. This heuristic validity criterion is actually too lax for magnetized plasma. A perturbation analysis of eq. (26) reveals that these higher-order gradient terms can be comparable to the zeroth-order terms (eq. (25)) even if $k_{\perp}L_n \gg 1$. Following some algebra, it is found that the two leading higher-order terms are linear in $\epsilon_{xy}$ and quadratic in $N_{||}$, respectively, such that the actual validity criterion for ray-tracing is $(|\epsilon_{xy}| + N_{||}^2) \frac{1}{k_{\perp} L_{n}} \ll 1$. (More details about this perturbation analysis can be found in Appendix A of \cite{biswas2020study}, although in that derivation $\nabla\bold{k}$ terms were erroneously neglected, which lead to the dropping of the $N_{||}^2$ term in the final ray-tracing validity criterion.) At initial launch of the LH wave $N_{||}^2 , |\epsilon_{xy}| \sim 1$, but they can both grow to be much larger as the ray continues to propagate. Specifically, $|\epsilon_{xy}|\propto n_e$, and so it rapidly increases as the ray propagates into the plasma. The restriction that $\epsilon_{xy}$ places on LH ray-tracing has been commented on before\cite{ott1979lower}. The following discussion is the first time it has been linked to asymmetric scattering in the context of LH waves.

\begin{figure}[!h]
\centering
\includegraphics[scale=0.6]{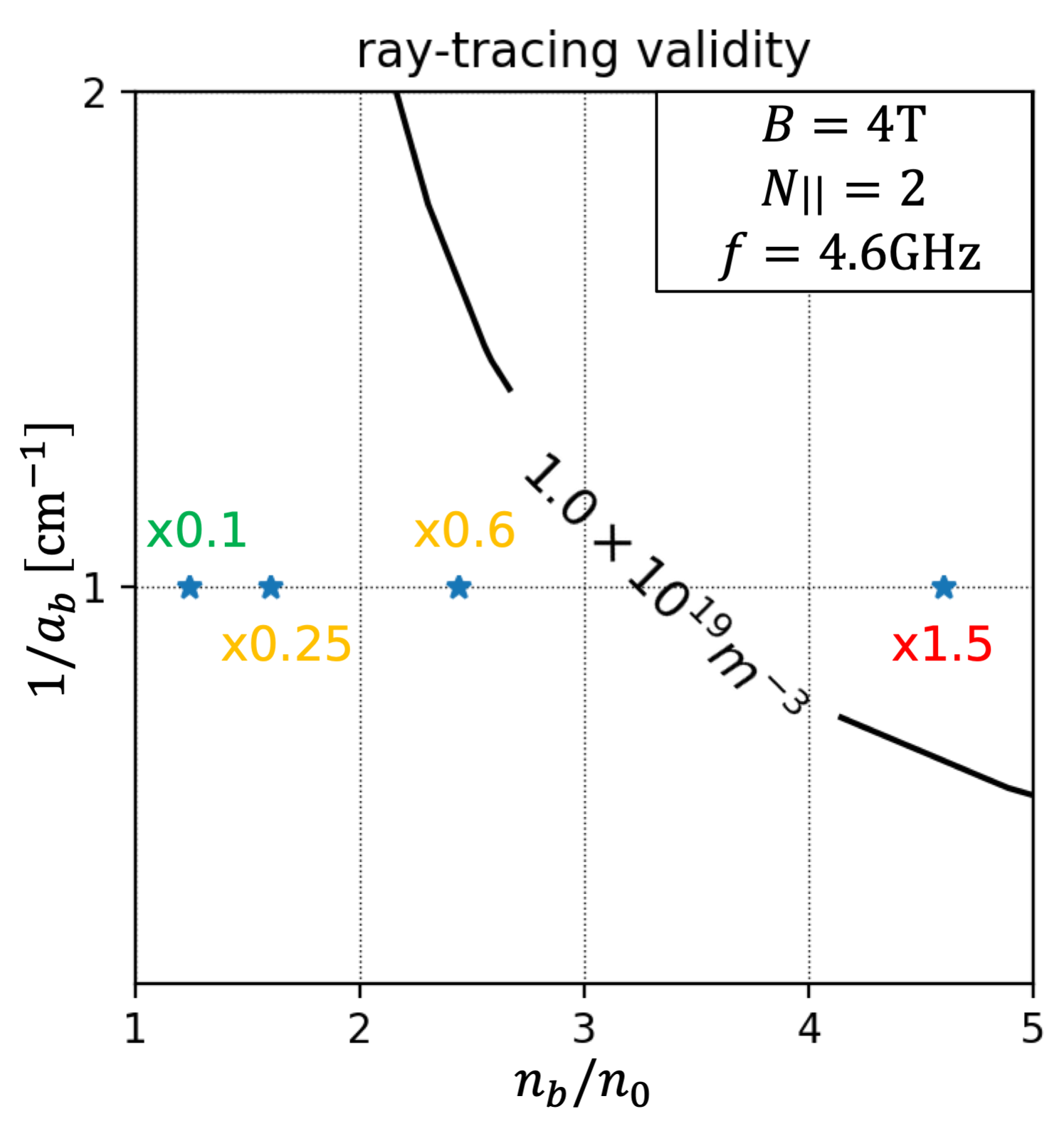}
\caption[font=5]{Ray-tracing validity for Gaussian filaments as a function of relative density ($n_b/n_0$) and radial width ($a_b$). Black line denotes validity limit at $n_0 = 1 \times 10^{19} \text{m}^{-3}$ (see \hyperref[sec:7.2]{Section 7.2} for more detail). Stars denote filaments that are used in scattering studies in Figures \ref{fig:5} and \ref{fig:6}. Numbers in color denote the ratio $n_b/n_{b,\text{max}}$.}
\label{fig:4}
\end{figure}

In deriving the new ray-tracing criterion for LH waves in a magnetized plasma, it was revealed that one of the leading higher-order gradient terms neglected in ray-tracing is linear in $\epsilon_{xy}$. This is precisely the term with information about the orientation of $\bold{B}$. In accordance, as $|\epsilon_{xy}|\frac{1}{k_{\perp} L_n}$ grows and becomes comparable to unity, asymmetric scattering also becomes important. This is shown numerically by simulating the scatter of LH waves from four increasingly dense Gaussian filaments. Figure \ref{fig:4} plots the validity regime of ray-tracing in the presence of a Gaussian filament as a function of $n_b/n_0$ and $a_b$. The incident wave is launched at 4.6$\,$GHz with $N_{||}=2$ and $\text{B} = 4\,\text{T}$. It is assumed that $L_{n}^{-1} \approx (\frac{n_b}{n_0}-1)/a_b$. The black line denotes the validity limit for $n_0 = 1 \times 10^{19} \text{m}^{-3}$. To the right of this line, $|\epsilon_{xy}|\frac{1}{k_{\perp} L_n} > 1$ and to the left $|\epsilon_{xy}|\frac{1}{k_{\perp} L_n} < 1$. (For simplicity, the $N_{||}^2$ term is ignored). The four starred points denote filaments with $a_b = 1$cm and $n_b/n_0 = [1.24, 1.6, 2.44, 4.6]$ (plotted left to right). Alternatively, these filaments satisfy $n_b = [0.1, 0.25, 0.6, 1.5]\times n_{b,\text{max}}$ where $n_{b,\text{max}}$ satisfies $|\epsilon_{xy}|\frac{1}{k_{\perp} L_n} = 1$. Qualitatively, the green point signifies a filament that is validly treated with ray-tracing since $|\epsilon_{xy}|\frac{1}{k_{\perp} L_n} \approx \frac{n_b}{n_{b,\text{max}}} = 0.1 < 1$. The yellow points are marginally valid, since $\frac{n_b}{n_{b,\text{max}}} < 1$ but also $\mathcal{O}(1)$. The red point, for which $\frac{n_b}{n_{b,\text{max}}} > 1$, certainly cannot be treated using ray-tracing.

\begin{figure}[!h]
\centering
\includegraphics[scale=0.7]{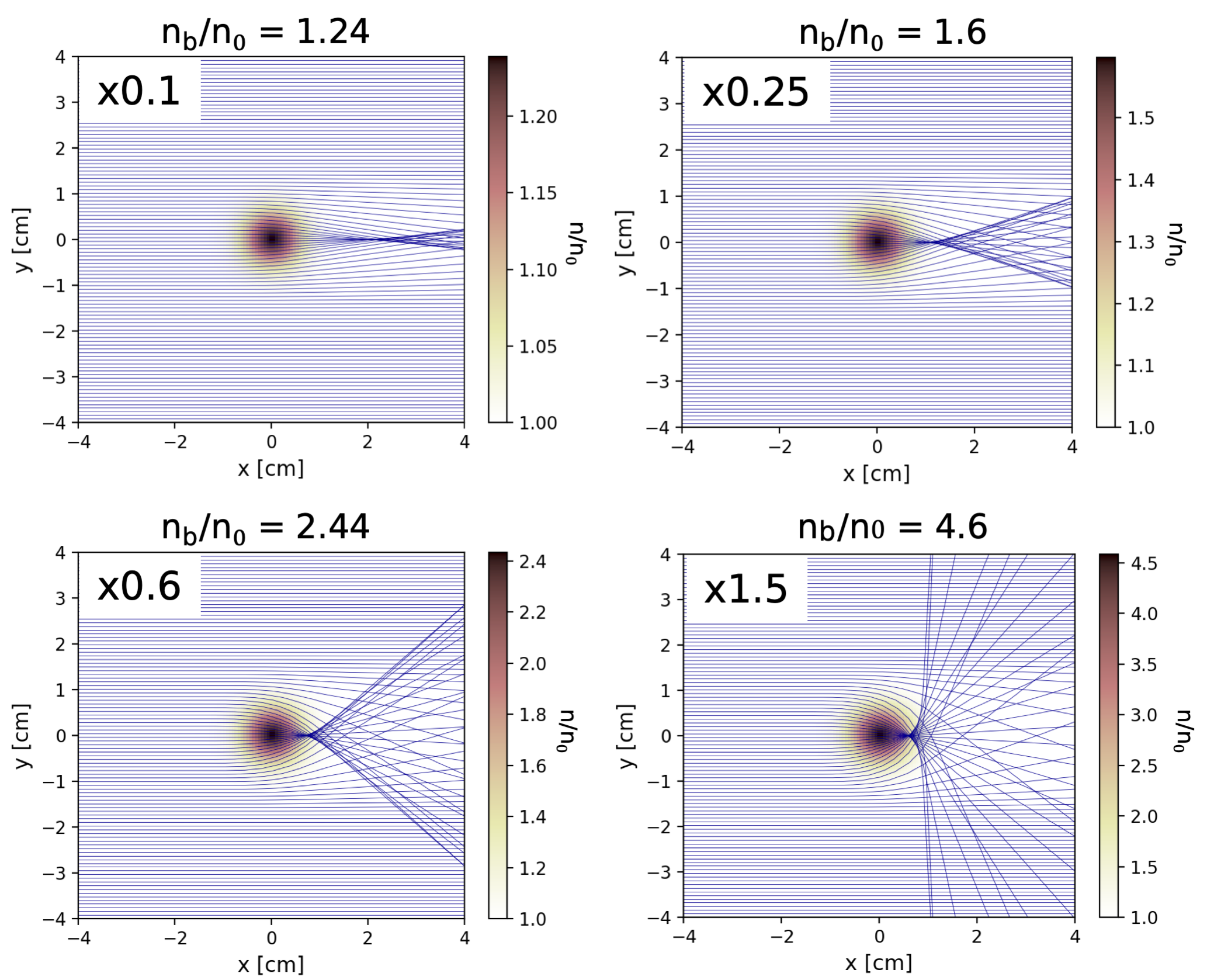}
\caption[font=5]{Ray-tracing simulations of LH waves scattering from a filament. f=4.6$\,$GHz, $N_{||}=2$, $\text{B}=4\,\text{T}$, and $n_0 = 1 \times 10^{19} \text{m}^{-3}$. Each subplot corresponds to a star on Figure \ref{fig:4}. Blue lines denote ray trajectories.}
\label{fig:5}
\end{figure}

Figure \ref{fig:5} plots the ray-trajectories of LH rays incident from the left and interacting with a filament. As $n_b/n_0$ increases, rays are more strongly refracted, resulting in a shadowing effect downstream of the filament. Notably, these ray-trajectories are always perfectly symmetric with respect to y=0.

\begin{figure}[h!]
\centering
\includegraphics[scale=0.7]{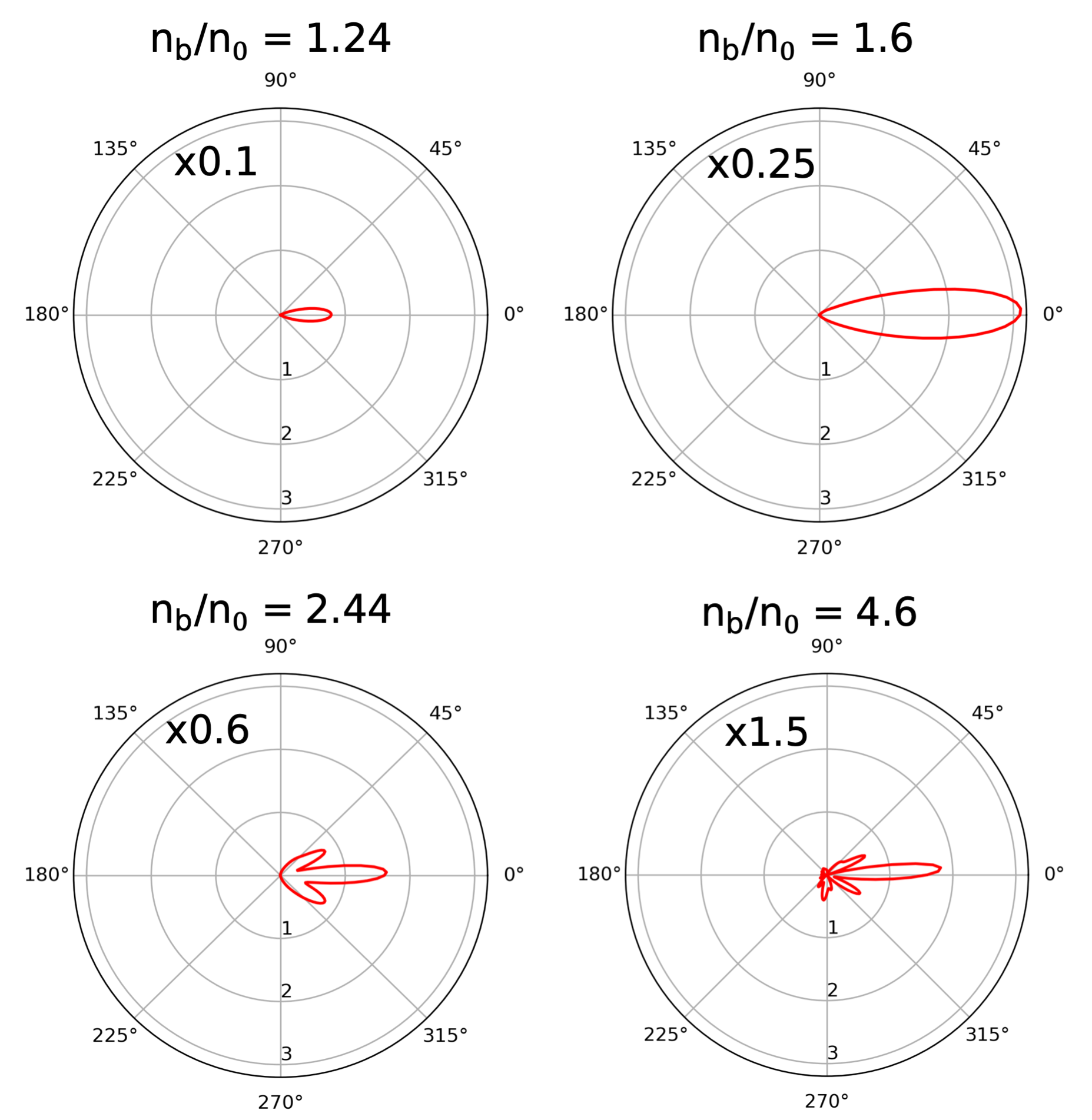}
\caption[font=5]{Polar plots of differential scattering-width $\sigma (\theta)$ calculated using the SAS method. Simulation parameters are same as in Figure \ref{fig:5}.}
\label{fig:6}
\end{figure}

In contrast, Figure \ref{fig:6} plots $\sigma (\theta)$ calculated using the SAS method for the same four cases simulated in Figure \ref{fig:5}. The SAS method (like all full-wave treatments) implicitly accounts for all terms in eq. (26). For $n_b/n_0=1.24$ and 1.6, $\sigma (\theta)$ is symmetric about $\theta = 0$. The profiles peak at $\theta =0$, signifying predominantly forward-scatter. At $n_b/n_0=2.44$, $\sigma (\theta)$ is slightly asymmetric because the side-lobe at $-40^{o}$ is larger than the one at $+40^{o}$. At $n_b/n_0=4.6$, $\sigma (\theta)$ is clearly asymmetric. Notably, the largest lobe is centered at $+5^{o}$. While a direct quantitative comparison between Figure \ref{fig:5} and \ref{fig:6} is not possible, it is clear that as $|\epsilon_{xy}|\frac{1}{k_{\perp} L}$ grows (filaments get denser), ray-tracing becomes less accurate because the increasingly important asymmetric scattering effect is ignored. It is important to note that filaments with $n_b/n_0 \gtrsim 2$ are common in the SOL\cite{zweben2002edge}, which signifies ray-tracing is inadequate for the treatment of LH wave scattering in realistic SOL turbulence.

\subsection{Modified wave-spectrum in front of LH antenna}
\label{sec:7.3}

\begin{figure}[!h]
\centering
\includegraphics[scale=0.6]{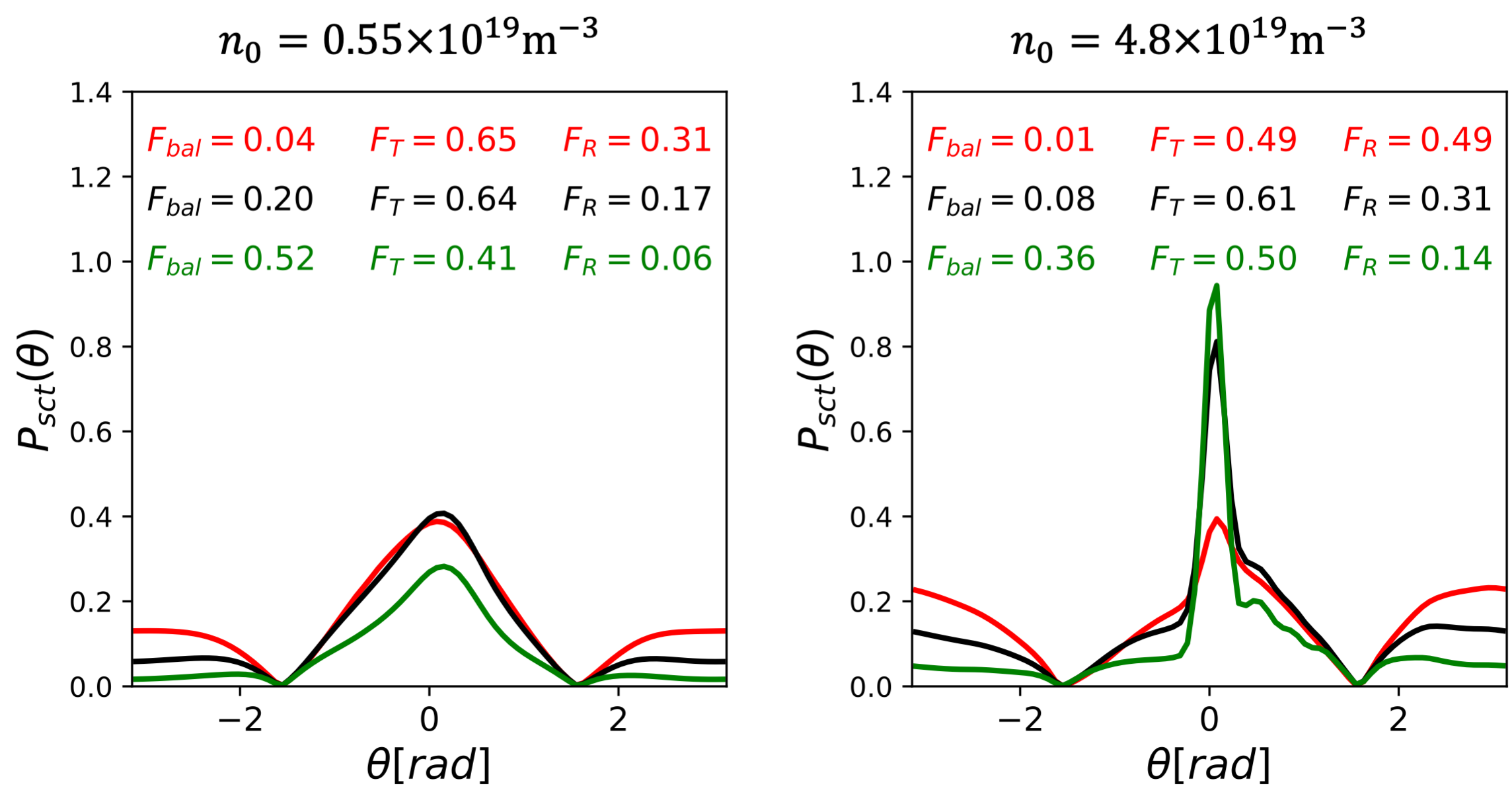}
\caption[font=5]{Modified LH wave-spectrum after interacting with turbulent slab. $|\theta| < \pi /2$ denotes transmitted power. $|\theta| > \pi /2$ denotes reflected power. Ballistic power not plotted. $f_p = $ 0.1 (green), 0.25 (black), 0.5 (red). Left and right plots assume low and high background density, respectively. $F_{\text{bal}}$, $F_{\text{T}}$, $F_{\text{R}}$ denote fractional ballistic, transmitted, and reflected power, respectively. $\sigma_{\text{eff}}(\theta)$ used is shown in Figure \ref{fig:8}.}
\label{fig:9}
\end{figure}

The incident wave parameters, background plasma parameters, and joint-PDF of filament parameters determine $\sigma_{\text{eff}}(\theta)$. The Markov Chain (MC) method is used to compute the modified wave-spectrum after the incident wave interacts with a slab layer of thickness $L_{x}$ and packing fraction $f_p$.  Figure \ref{fig:9} plots the modified wave-spectrums resulting from the $\sigma_{\text{eff}}(\theta)$ shown in Figure \ref{fig:8}. $L_{x} = 2.5\,$cm, which is the typical gap between the LH antenna and the separatrix in C-Mod. Green, black, and red lines denote $f_p=[0.1,0.25,0.5]$. Note that the ballistic power fraction is not plotted (if it were, it would be a Dirac-delta plotted at $\theta = 0$). In the low-density ($n_0 = 0.55 \times 10^{19} \text{m}^{-3}$) case, the modified wave-spectrum is smoothly broadened in $\theta$-space, with a peak centered at $\theta = 0$. Increase in $f_p$ leads to a decrease in ballistic power, and increase in reflected power. This is expected, since $\Sigma_{\text{eff}}$, the inverse mean-free-path to scatter, is linearly proportional to $f_p$. In the high-density ($n_0 = 4.8 \times 10^{19} \text{m}^{-3}$) case, the modified wave-spectrum is significantly asymmetric, with net power scattered in the +$\theta$-direction. Naturally, this is the result of $\sigma_{\text{eff}}(\theta)$ in the high-density case being very asymmetric. Again, ballistic power decreases and reflected power increases with $f_p$. In comparing the low and high-density cases, it is found that the high-density case results in significantly greater reflected power. This is due to $\sigma_{\text{eff}}$ being larger in the high-density cases, as can be seen by inspecting Figure \ref{fig:8}.

\subsection{SAS-MC compared with ray-tracing}
\label{sec:7.4}

A comparison study between the SAS-MC model and ray-tracing model is conducted. Rays are launched in a slab geometry, and are incident normal to a slab comprised of randomly generated filaments (see \hyperref[sec:6.2]{Section 6.2}). A ray terminates when it leaves the slab (either reflected backward or transmitted forward), at which point the angle between the ray's perpendicular group velocity ($\bold{v}_{gr_{\perp}}$) and $\hat{e}_x$ is tallied. This poloidal angle is the direction that the ray continues to propagate and radiate power away from the slab. It is therefore equivalent to $\theta$ in the SAS-MC model. Following multiple ray launches, a histogram of these tallies is constructed. This histogram, once properly normalized, is equivalent to a modified wave-spectrum that can be compared with the wave-spectrum computed with the SAS-MC model.

\begin{figure}[!h]
\centering
\includegraphics[width=17cm, height=5cm]{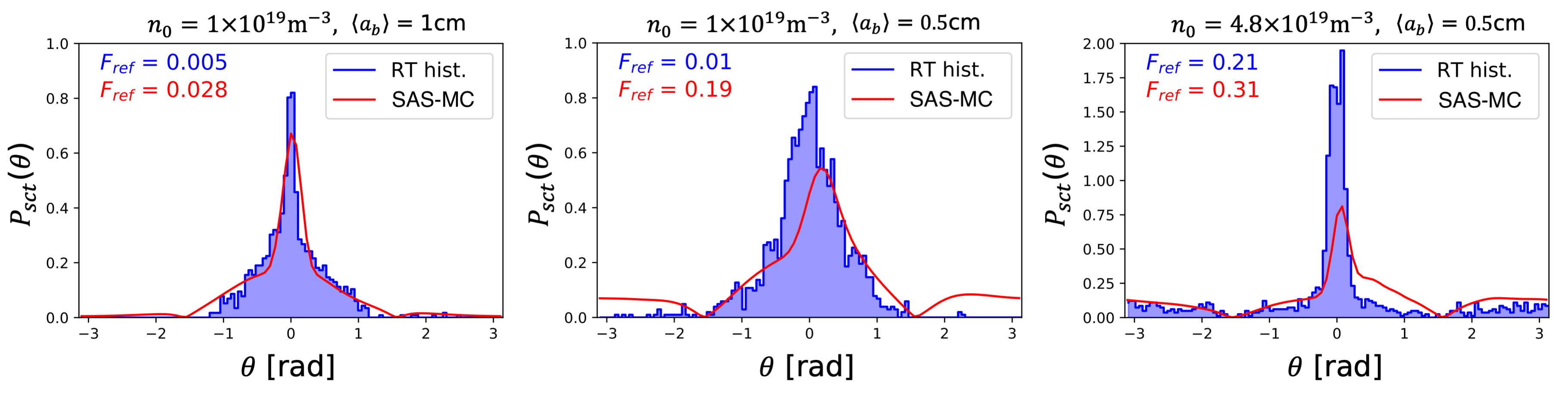}
\caption[font=5]{Comparison between SAS-MC and ray-tracing slab model. Blue bars denote histogram of ray $\bold{v}_{gr_{\perp}}$ angle after leaving slab. Red line denotes modified wave-spectrum $P_{sct} (\theta)$ calculated using SAS-MC method. $L_x = 2.5\,$cm. $f_p = 0.25$. Filament joint-PDF parameters are same as in Figure \ref{fig:7} unless otherwise noted in subplot title. $F_{\text{ref}}$ denotes fractional power reflected in (blue) ray-tracing and (red) SAS-MC model. Ballistic power and un-scattered rays are not plotted.}
\label{fig:10}
\end{figure}

Figure \ref{fig:10} plots modified wave-spectrums computed using the SAS-MC model and the ray-tracing model for statistically identical turbulent slabs. Three different cases are ran. All cases assume LH rays incident at 4.6GHz and $N_{||}=2$. $\text{B}=4\,\text{T}$ and $L_x=2.5\,$cm. In the first case, $n_0 = 1 \times 10^{19}\,\text{m}^{-3}$, and the joint-PDF in Figure \ref{fig:7} is assumed. The SAS-MC model and ray-tracing model show good agreement in wave-spectrum. Both predict low reflected power fractions. The wave-spectrum computed with the SAS-MC model is fairly symmetric. This means that asymmetric scatter was weak, and therefore the ray-tracing approximation was valid. Thus, the good agreement between the two models. In the next case, $\langle a_b \rangle$ is halved to 0.5 cm. The SAS-MC model results in an asymmetric wave-spectrum, such that the peak is shifted to +0.2rad. The onset of significant asymmetric scattering is caused the decrease in $L_{n}$. At the same time, the ray-tracing approximation begins to break down. As a result, the ray-tracing results (which are symmetric) begin to deviate from the SAS-MC results. Notably, the ray-tracing model severely under-predicts the fraction of reflected power compared to the SAS-MC model. The last case increases background density to $n_0 = 4.8 \times 10^{19}\,\text{m}^{-3}$. Now, $\epsilon_{xy}$ is large enough that asymmetric scattering is quite strong and the ray-tracing approximation is surely invalid. As a result, the two models result in very different wave-spectrums. Ray-tracing predicts a scattered wave-spectrum with a large central peak between $\pm 0.2\,$rad. In contrast, the SAS-MC model predicts a smaller central peak slightly shifted in the +$\theta$-direction. The tail to the right of this peak is significantly larger than the one on the left. Lastly, SAS-MC model results in $\sim$50\% more power being reflected than the ray-tracing model.

\section{Impact of scattering on LHCD}
\label{sec:8}

In typical ray-tracing/Fokker-Planck simulations, the initial perpendicular group-velocity $\bold{v}_{gr\perp}\equiv \partial \omega /\partial \bold{k}_{\perp}$ is assumed co-parallel with the unit vector normal to the flux surface ($\hat{e}_{\nabla \psi}$). Therefore, the angle between these two vectors, $\chi \equiv \angle (\bold{v}_{gr\perp}, \hat{e}_{\nabla \psi})$, is usually zero. Scattering caused by edge density fluctuations can rotate $\bold{v}_{gr\perp}$ leading to a broadened wave-spectrum in $\chi$-space, as evidenced by LH electric field vector measurements in C-Mod \cite{martin2019experimental}. This rotation can modify the ray-path so that single-pass damping is strengthen or weakened, depending on the sign of $\chi$ \cite{baek2020role}.

The $\chi$ angle in the tokamak frame and $\theta$ in the slab geometry (as defined in Figure \ref{fig:1}) are identical. However, the orientation of $\chi$ (i.e. whether positive $\chi$ points upwards or downwards in the tokamak frame) depends on the sign of $N_{||}$, and the orientation of toroidal magnetic field and current in the tokamak. Given that $N_{||}$ must always be directed opposite to the plasma current (in order to drive co-current via electron Landau damping), it is found that $\chi$ (or equivalently $\theta$) must be oriented such that $\chi > 0$ rotates the ray trajectory away from the core. Conversely, $\chi < 0$ rotates the ray towards the core. This is true in all tokamak orientations.

\begin{figure}[!h]
\centering
\includegraphics[scale=0.4]{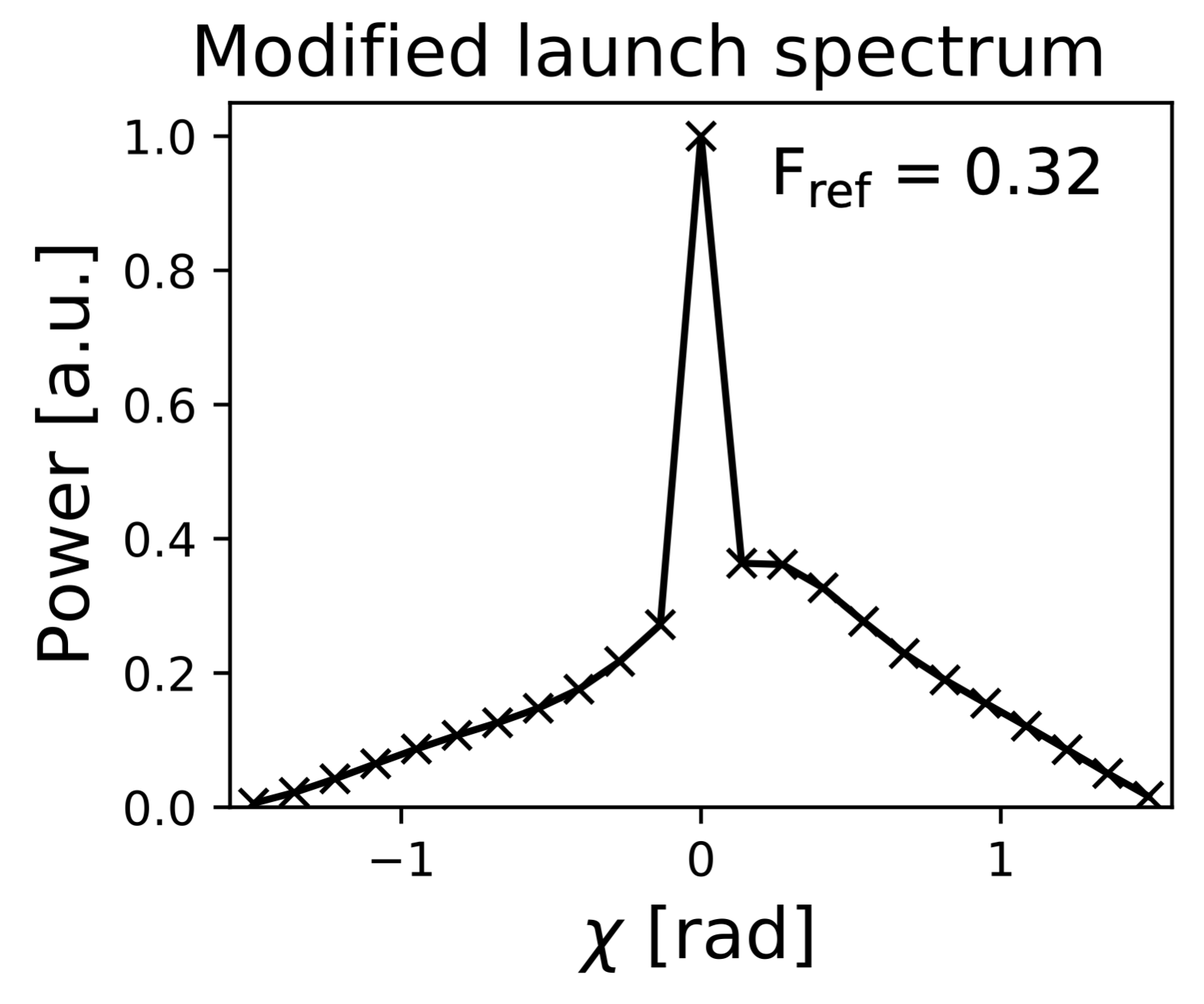}
\caption[font=5]{Modified wave-spectrum launched in GENRAY/CQL3D simulation of LHCD in Alcator C-Mod. Wave-spectrum calculated using SAS-MC model for slow-wave launched at 4.6$\,$GHz and $N_{||}=1.6$. SOL background density $n_0 = 1 \times 10^{19}\,\text{m}^{-3}$, packing-fraction $f_p = 0.25$, and SOL width $L_x = 2.5\,$cm is assumed. Filament joint-PDF parameters are same as in Figure \ref{fig:7}. Crosses show discretization of wave-spectrum into rays for use in GENRAY. Spike at $\chi =0$ is due to ballistic power. Reflected power is ignored.}
\label{fig:12}
\end{figure}

Thus, the transmitted wave-spectrum calculated using the SAS-MC model can be coupled to GENRAY/CQL3D to study its impact on LHCD. A well-studied \cite{mumgaard2015lower}, low-density, L-mode discharge is modeled. This upper single-null discharge, with $\overline{n}_e = 0.52 \times 10^{20}\,\text{m}^{-3}$, $I_p = 530\,\textrm{kA}$, and $\text{B}=5.4\,$T, achieves non-inductive current drive using 850$\,$kW of LH power launched at 4.6$\,$GHz with $N_{||} = -1.6$. ($\overline{n}_e$ is line-averaged electron density and $I_p$ is plasma current.) It is assumed that $85\%$ of power is coupled to the primary lobe. In GENRAY, this primary lobe is centered at $N_{||}=-1.6$, and is discretized into 12 bins in $N_{||}$-space. Each bin is further discretized into 23 rays to model wave-spectrum broadening in $\chi$-space. Figure \ref{fig:12} plots the SAS-MC calculated transmitted wave-spectrum assuming SOL background density $n_0 = 1 \times 10^{19}\,\text{m}^{-3}$, SOL width $L_x = 2.5\,$cm, and packing fraction $f_{p}=0.25$. Filament joint-PDF is the same as in Figure \ref{fig:7}. The spike at $\chi$=0 accounts for ballistic power. Note that the reflected power, which accounts for roughly $30\%$ of power in the primary lobe, is assumed lost and therefore not modeled in GENRAY. Lastly, the spatial height of the launcher is modeled as 4 poloidal points in the outer mid-plane. In total, 1104 rays are launched to ensure a converged solution.

The rays are launched from the separatrix, but are allowed to propagate into the SOL after first-pass.  Here, the rays will either reflect at the cutoff density or specularly reflect from the vessel wall back towards the core. Due to low temperatures in the SOL, collisional damping in non-negligible. It is found that the SOL topology and presence of a divertor can significantly affect the calculated core CD profiles. Therefore, the two-point model is used to accurately generate the SOL\cite{shiraiwa2015impact}. Once the proper SOL geometry is set, parameters like SOL e-folding width and divertor temperature do not strongly affect core CD results for cases with the $\chi$-broadened wave-spectrum. 

\begin{figure}[!h]
\centering
\includegraphics[scale=0.8]{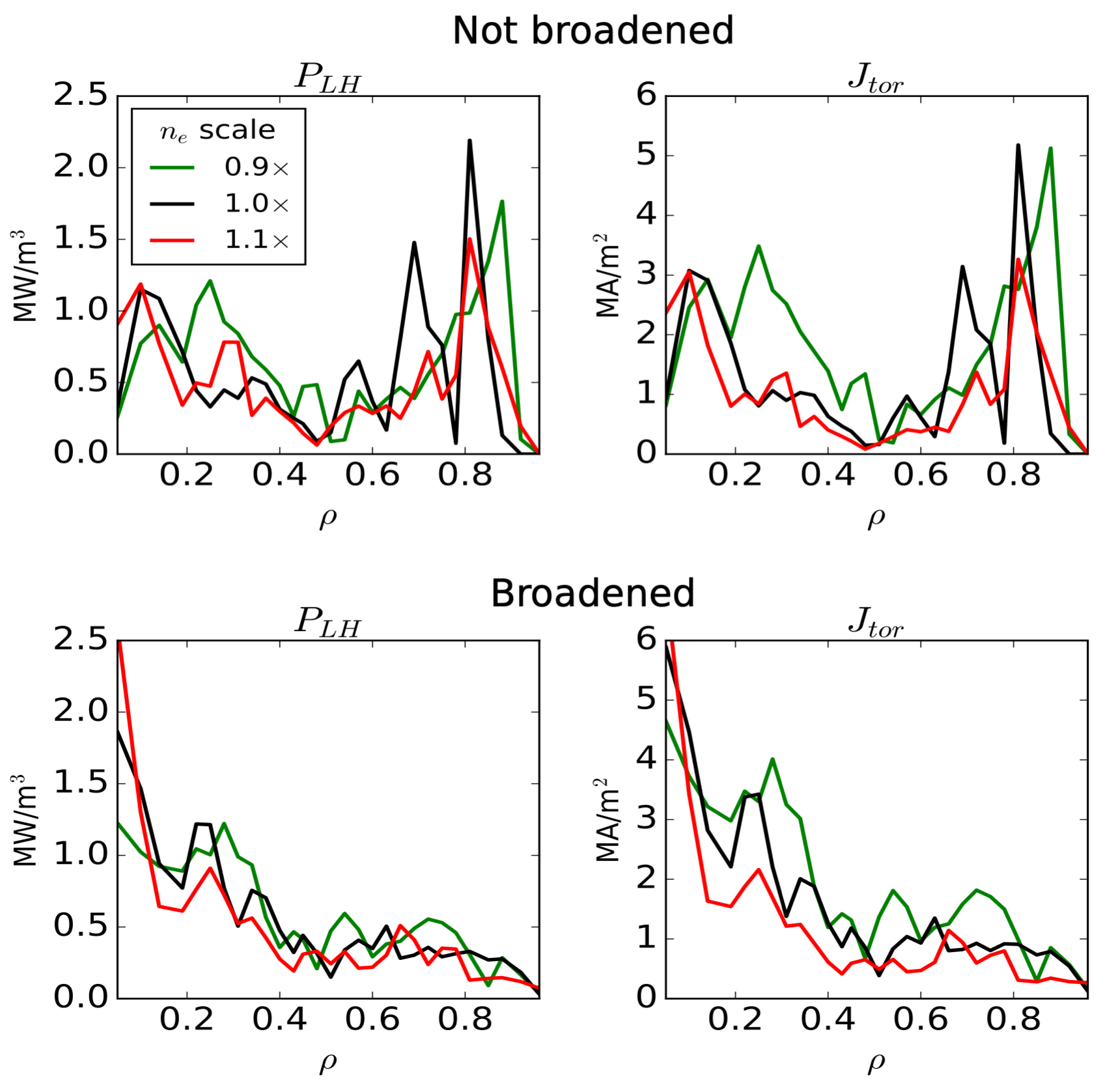}
\caption[font=5]{Core LH power deposition and driven current density profiles in C-Mod L-mode discharge \#1101104011 at t=1.10s, modeled with GENRAY/CQL3D. $\overline{n_e} = 0.52\times 10^{20} \,\text{m}^{-3}$, $I_p = 530\,\text{kA}$, and $\text{B}=5.4\,$T. 850$\,$kW of LH power launched at 4.6$\,$GHz and $N_{||}=1.6$. (Top) simulations with no wave-spectrum broadening. (Bottom) simulations with broadened angular wave-spectrum shown in Figure \ref{fig:12}. Green, black, red lines denote core background density is scaled x0.9, x1.0, x1.1 the nominal value, respectively.}
\label{fig:13}
\end{figure}

Figure \ref{fig:13} plots the calculated core power deposition and CD profile in this C-Mod discharge. The core density is scaled $\pm 10 \%$ to assess the sensitivity of these results. The top figures, for which the launched wave-spectrum was not broadened in $\chi$-space, reveal core profiles that are robustly peaked at $\rho \approx 0.8$. A smaller peak exists on-axis, though it shifts to $\rho \approx 0.25$ when the background density is decreased $-10\%$. A robust current valley exists at $\rho \approx 0.5$. The bottom plots model a launched wave-spectrum that is $\chi$-broadened. These profiles are remarkably different from the cases without broadening. There is a $65\%$ increase in power deposited near-axis ($\rho < 0.5$), leading to profiles that are robustly peaked on-axis. There are also no large off-axis peaks.

\begin{figure}[!h]
\centering
\includegraphics[scale=0.6]{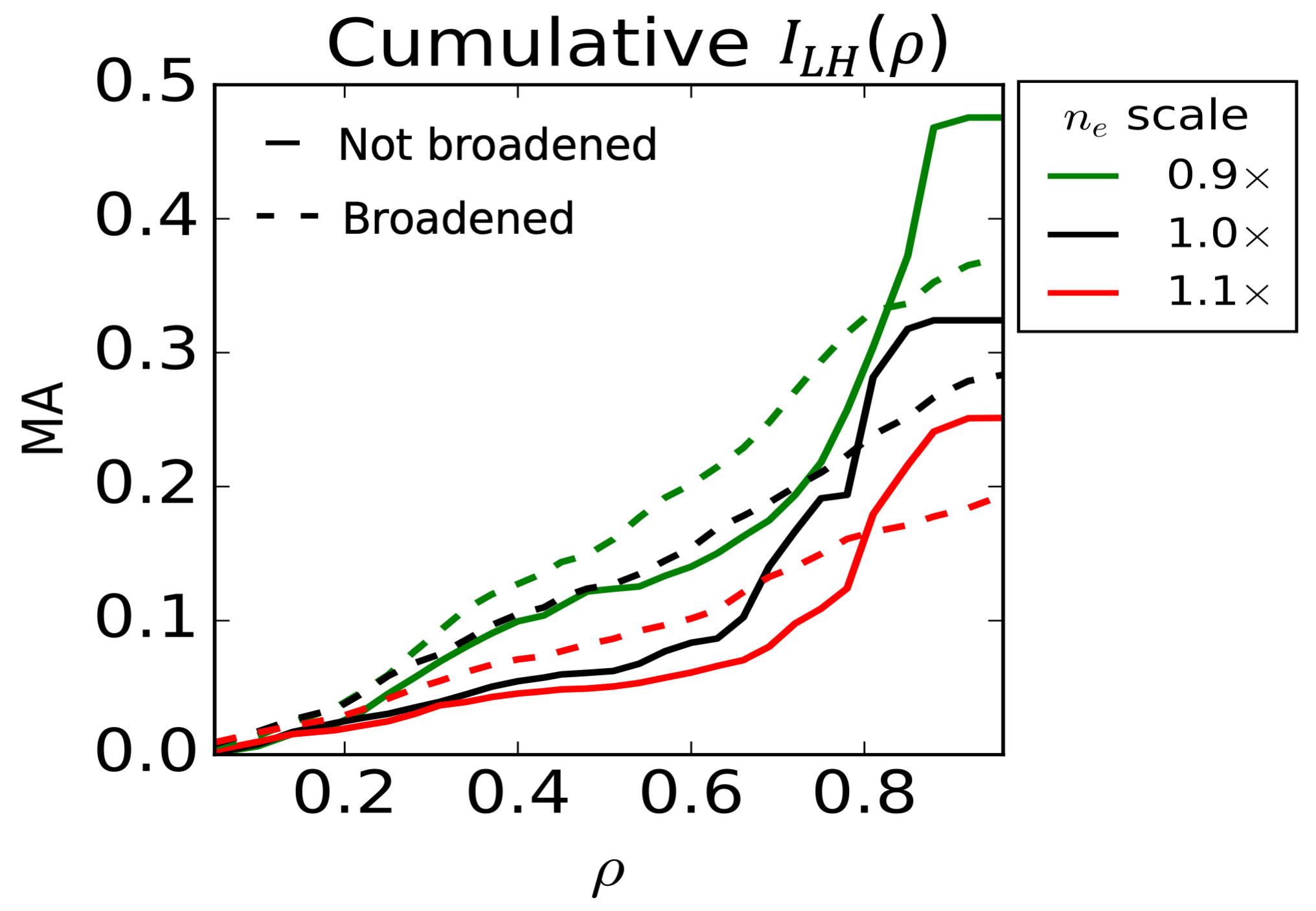}
\caption[font=5]{Cumulative core LH current driven, modeled in GENRAY/CQL3D. Simulation parameters same as in Figure \ref{fig:13}.}
\label{fig:14}
\end{figure}

Figure \ref{fig:14} plots the cumulative CD profile. The $\chi$-broadened cases result in roughly linear profiles and greater current driven near-axis. In contrast, the cases without broadening result in CD preferentially in the off-axis ($\rho > 0.7$) region.  The total LH current is $10-20\%$ lower in the broadened cases. This is partly due to $\sim\!30\%$ of incident power being reflected in the SAS-MC model, and therefore not being launched in GENRAY for the $\chi$-broadened cases.

\begin{figure}[!h]
\centering
\includegraphics[scale=0.7]{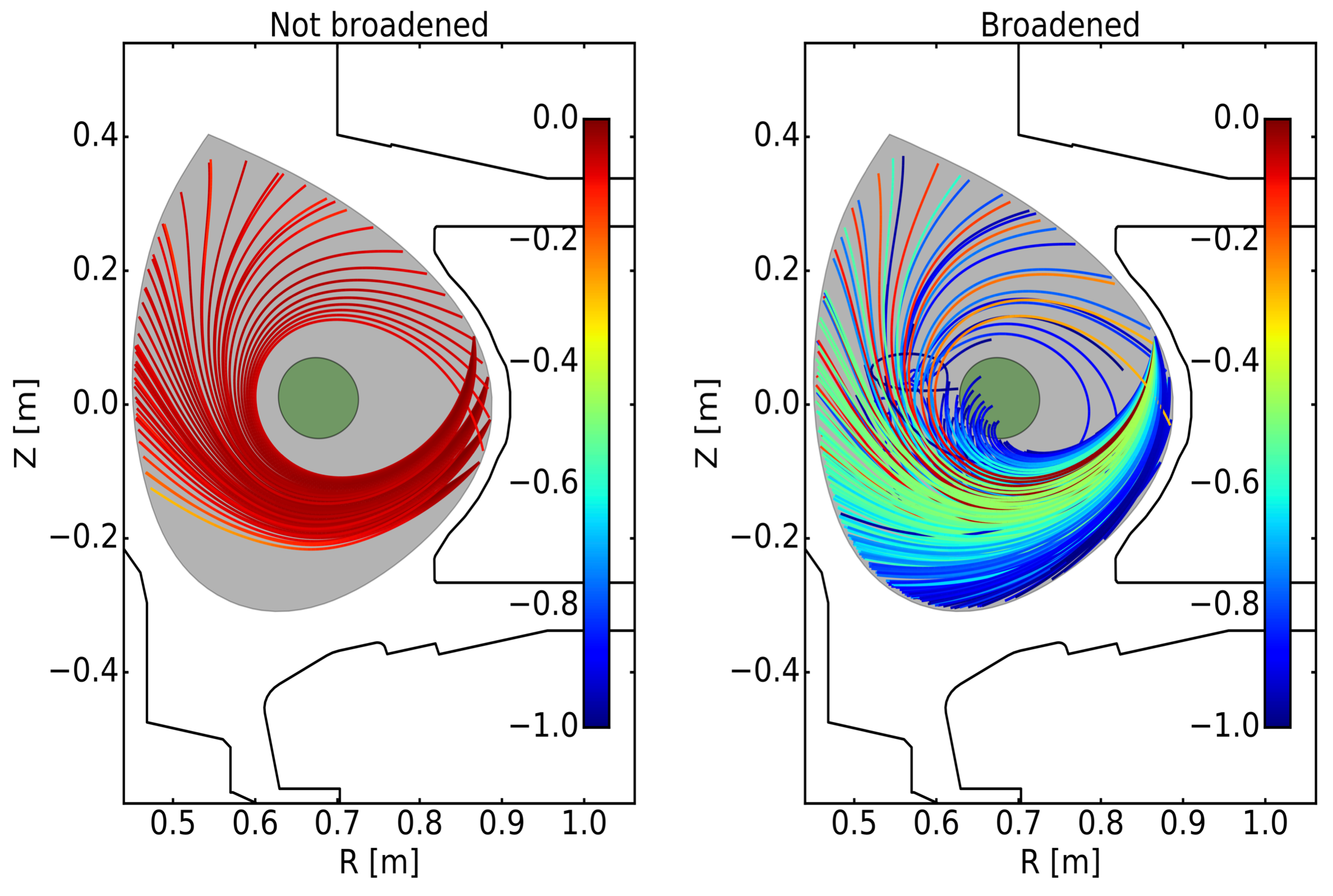}
\caption[font=5]{Poloidal projection of first-pass ray-trajectories in C-Mod discharge. Simulation parameters are the same as in Figure \ref{fig:13} for x1.0 scaled density case. Colored lines denote ray-trajectories. The color of lines denote $\text{log}_{10}$ power in ray, normalized to initial power in highest-power ray. Gray patch denotes core region ($\rho < 1$). Green patch denotes near-axis region ($\rho < 0.2$).}
\label{fig:15}
\end{figure}

Figure \ref{fig:15} plots the ray-trajectories during first-pass. Ray color denotes the logarithmic power in the ray, normalized to initial power in the highest-powered ray. In the case with no broadening, rays cannot propagate to the hot magnetic axis, and therefore cannot Landau damp strongly. In contrast, $\chi$-broadening ``fans'' out the initial ray trajectories. Notably, rays that are sufficiently rotated inwards ($\chi < 0$) strongly Landau damp in the hot near-axis plasma. Even though this is a small fraction of the incident power, it is sufficient to seed a supra-thermal electron tail near-axis. As a result, additional rays can quasi-linearly damp on this tail on subsequent passes through the core. In the case without broadening, there is insufficient on-axis power for this seeding effect. As a result, on-axis current drive is relatively low.

Note that the modified wave-spectrum has a net effect of deflecting power away from the core on first pass. Paradoxically, near-axis CD increases. Again, this is attributed to the small fraction of power deflected inwards that seeds a near-axis supra-thermal electron tail. It is possible this phenomenon does not extend to high-density discharges, where stronger asymmetric scattering will deflect a greater fraction of power outwards.

\section{Conclusion}
\label{sec:9}
A hybrid Semi-Analytic Scattering Markov chain (SAS-MC) model is formulated to calculate the modified wave-spectrum of an RF wave propagating through a turbulent SOL. First, a semi-analytic full-wave technique is adopted to calculate the scattered power from a SOL filament. This technique is generalized to account for filaments with radially-varying densities. Next, an effective differential scattering-width is derived for a statistical ensemble of filaments. Lastly, the SOL is modeled as a slab, and the modified wave-spectrum is found by solving the radiative transfer equation using a Markov chain technique. This model is applied to the case of Lower Hybrid launch for driving current in a tokamak. GENRAY/CQL3D is used to model the impact of the modified wave-spectrum on current drive in Alcator C-Mod.

In calculating the differential scattering-width, it is found that the scattered power can be asymmetrically directed (in the y-direction). This is true even for the effective scattering-width, which averages over the statistical properties of filaments. Previous RF scattering models have either used the drift-wave approximation and/or the ray-tracing approximation. As a result, they fail to account for this important asymmetric effect.

The SAS-MC model is compared to the ray-tracing treatment of LH wave scattering. By retaining full-wave effects, the SAS-MC model is able to produce a significantly asymmetric transmitted wave-spectrum. As stated previously, ray-tracing cannot replicate this effect. 

The SAS-MC model is compared to PETRA-M, which self-consistently models full-wave interactions in the presence of multiple filaments. Both models predict $F_{\text{ref}}$ increases with $\langle n_b/n_0 \rangle$, $\langle a_b \rangle^{-1}$, and $L_{x}$, as do previous analytic scattering models. Assuming a low background density, and realistic SOL packing-fraction and width, the two models agree in the calculated $F_{\text{ref}}$. As packing-fraction rises, the SAS-MC model increasingly over-predicts $F_{\text{ref}}$, which suggests this is a result of the far-field approximation breaking down. Nevertheless, the SAS-MC model retains full-wave effects for scattering from a single filament, and is therefore a significant improvement over previous reduced models for scattering.

A modified wave-spectrum is calculated for LH launch in a low-$\bar{n}_{e}$ Alcator C-Mod discharge. Roughly $30\%$ of launched power is reflected back into the SOL. The transmitted wave-spectrum is coupled to GENRAY/CQL3D, resulting in a significantly altered core CD profile. Notably, the on-axis current is increased, and off-axis peaks are greatly mitigated. This is attributed to a portion of the modified wave-spectrum that is rotated such that it damps on-axis during first-pass. This seeds a supra-thermal electron population on which rays preferentially Landau damp during subsequent passes through the core. The result is a CD profile that better matches experimental measurements in low-$\bar{n}_{e}$ discharges\cite{mumgaard2015lower}, which robustly feature monotonic profiles that peak on-axis.

The asymmetric scattering effect is stronger at high SOL densities, and result in a significant net deflection of launched LH power away from the core. This may induce greater parasitic losses in the edge, either through collisional damping or PDI. This warrants the investigation of asymmetric scattering as a possible explanation to the LHCD density limit \cite{wallace2010absorption}.

Lastly, it should be noted that the SAS-MC model is not limited to the LH frequency range. For example, this model is well-suited for the study of wave-spectrum broadening of the electron-cyclotron wave in the tokamak SOL. The relatively larger $k_{\perp}$ of the electron-cyclotron wave means the $k_{\perp}d\gg1$ criterion for the far-field approximation is more strongly satisfied than in the case of LH waves.

\newpage
\begin{appendix}
\addcontentsline{toc}{section}{Appendices}
\section*{Appendices}
\setcounter{equation}{0}
\renewcommand{\theequation}{A.\arabic{equation}}
\renewcommand{\thesubsection}{}

\subsection{Appendix A: Electric field in cylindrical coordinates}
\label{sec:A1}

The incident plane-wave is assumed to be have a wave-vector $\bold{k} = k_{\perp}\hat{e}_{x} + k_{||}\hat{e}_{z}$. Given the background is homogeneous, the incident wave solution is 
\begin{equation}
\bold{E}_{0} = (\xi_{0x}\hat{e}_{x}+\xi_{0y}\hat{e}_{y}+\xi_{0z}\hat{e}_{z})e^{i (k_{\perp}x + k_{||}z-\omega t)}
\end{equation}
$\bar{\xi}_0 = \{\xi_{0x},\xi_{0y},\xi_{0z}\}$ is the wave polarization. It can be evaluated by finding the null-space of the dispersion tensor for the given frequency and incident wave-vector.
The following transformation to cylindrical coordinates is used:
\begin{subequations}
\begin{align}
\hat{e}_x & = \hat{e}_{\rho}\cos{\theta} - \hat{e}_{\theta}\sin{\theta}\\
\hat{e}_y & = \hat{e}_{\rho}\sin{\theta} + \hat{e}_{\theta}\cos{\theta}\\
\hat{e}_z & = \hat{e}_z
\end{align}
\end{subequations}
to yield
\begin{equation}
\bold{E}_{0} = \left[\hat{e}_{\rho}(\xi_{0x}\cos{\theta}+\xi_{0y}\sin{\theta}) +\hat{e}_{\theta}(-\xi_{0x}\sin{\theta}+\xi_{0y}\cos{\theta})+\hat{e}_{z}\xi_{0z}\right]e^{i (k_{\perp}x + k_{||}z-\omega t)}
\end{equation}
Next, eq. (A.3) and the Jacobi-Anger identity are employed to cast the incident wave as a series solution in cylindrical coordinates. This results in eqs. (1) and (2). Equation (1) can be generalized to the non-incident waves for the following reason. The plane-wave $\bold{E}_0$, as formulated in eq. (1), is the known solution to this equation if the correct values of $k_{\perp}$ and $\bar{\xi}_0$ are used. In addition, each poloidal mode-number term in the series is a solution to the wave equation. It therefore follows that eqs. (1) can describe all other waves ($j \neq 0$) given the appropriate coefficients $E_{jm}$ are found.

\subsection{Appendix B: ``Flat top'' filament system of equations}
\label{sec:A2}

A system of equations must be formulated to find $E_{jm}$ for $j=1,...,4$.  $j=0$ denotes the incident wave. $j=1,2$ denote the slow, fast waves in the filament. $j=3,4$ denote the slow, fast scattered waves outside the filament. Assuming no free charge or current on the cylinder edge, the following boundary conditions are imposed:
\begin{subequations}
\begin{align}
\hat{e}_{\rho} \cdot (\bold{D}_{0}+\bold{D}_{1}+\bold{D}_{2})|_{\rho = a_{b}} & = \hat{e}_{\rho} \cdot (\bold{D}_{3}+\bold{D}_{4})|_{\rho = a_{b}} \\
\hat{e}_{\rho} \cdot (\bold{B}_{0}+\bold{B}_{1}+\bold{B}_{2})|_{\rho = a_{b}} & = \hat{e}_{\rho} \cdot (\bold{B}_{3}+\bold{B}_{4})|_{\rho = a_{b}} \\
\hat{e}_{\rho} \cross (\bold{E}_{0}+\bold{E}_{1}+\bold{E}_{2})|_{\rho = a_{b}} & = \hat{e}_{\rho} \cross (\bold{E}_{3}+\bold{E}_{4})|_{\rho = a_{b}} \\
\hat{e}_{\rho} \cross (\bold{B}_{0}+\bold{B}_{1}+\bold{B}_{2})|_{\rho = a_{b}} & = \hat{e}_{\rho} \cross (\bold{B}_{3}+\bold{B}_{4})|_{\rho = a_{b}}
\end{align}
\end{subequations}
where $\bold{D}_{j}$ is the electric displacement field of wave $j$. Equations (A.4) provide six constraints, but only four are independent. Myra and D'Ippolito (2010) \cite{myra2010scattering} employ eqs. (A.4a,c) and require $B_{z}$ to be continuous at the boundary. This paper follows this prescription.

The field solution have the poloidal dependence $e^{i m \theta}$. These exponential terms are orthogonal, and therefore the $m$-th terms must independently satisfy the boundary conditions. The following quantities are introduced:
\begin{subequations}
\begin{align}
\mathcal{D}_{jm} & = \epsilon_{\perp} W_{j\rho m} - i \epsilon_{xy} W_{j\theta m}\\
M_{jm} & = \xi_{jy}\left(k_{j \perp} J_{m}^{''} + \frac{1}{\rho}J_{m}^{'}-\frac{m^2}{k_{j \perp} \rho^{2}}J_{m}\right)
\end{align}
\end{subequations}
where $\epsilon_{\perp}$ and $\epsilon_{xy}$ are components of the dielectric tensor in the Stix frame\cite{stix1992waves}. The argument of $J_{m}(k_{j \perp} \rho)$ has been suppressed. Again, $J$ must be replaced with the appropriate type of Bessel/Hankel function for the wave. $E_{j}\mathcal{D}_{jm}$ is proportional to $\bold{D}_{jm}\cdot \hat{e}_{\rho}$. $E_{j}M_{jm}$ is proportional to $\bold{B}_{jm}\cdot \hat{e}_{z}$. Equations (A.4) and (A.5) are used to formulate the following linear system of equations:
\begin{equation}
\begin{bmatrix}
W_{1\theta m} & W_{2\theta m} & -W_{3\theta m} &-W_{4\theta m}\\
M_{1m} & M_{2 m} & -M_{3m} &-M_{4m}\\
\mathcal{D}_{1m} & \mathcal{D}_{2 m} & -\mathcal{D}_{3m} &-\mathcal{D}_{4m}\\
W_{1z m} & W_{2z m} & -W_{3z m} &-W_{4z m}\\
\end{bmatrix}
\cdot
\begin{bmatrix}
E_{1m}\\
E_{2m}\\
E_{3m}\\
E_{4m}
\end{bmatrix}
= -E_{0m}
\begin{bmatrix}
W_{0\theta m}\\
M_{0m}\\
\mathcal{D}_{0m}\\
W_{0zm}
\end{bmatrix}
\end{equation}
which is evaluated at $\rho = a_b$. The only unknown is the column vector on the LHS. It is solved for by inverting the 4x4 matrix. This process is repeated for each poloidal mode-number.

\subsection{Appendix C: Radially in-homogeneous filament system of equations}
\label{sec:A3}
In general, there are $4(R+1)$ unknown wave coefficients and $4(R+1)$ independent boundary equations, making this problem solvable for any $R$. For convenience, the wave indices are reordered. Waves $j=0,1$ denote the slow, fast wave (respectively) in the inner-most ($r=0$) bin. Waves $j=4r-2,...,4r+1$ are the slow $H_{m}^1$, slow $H_{m}^2$, fast $H_{m}^1$, and fast $H_{m}^2$ contributions (respectively) in bin $r>0$. Waves $j=4R+2,4R+3$ are the slow, fast scattered waves outside the cylinder. Lastly, wave $j=4R+4$ is the incident wave.
The first four matching relations (with the $m$ subscript suppressed) are:
\begin{equation}
\begin{bmatrix}
-W_{0\theta} & -W_{1\theta} &W_{2\theta} & W_{3\theta}& W_{4\theta} &W_{5\theta}\\
-M_{0} & -M_{1}& M_{2} & M_{3} &M_{4} &M_{5}\\
-\mathcal{D}_{0} & -\mathcal{D}_{1}& \mathcal{D}_{2} & \mathcal{D}_{3} &\mathcal{D}_{4} &\mathcal{D}_{5}\\
-W_{0z} & -W_{1z} &W_{2z} & W_{3z}& W_{4z} &W_{5z}
\end{bmatrix}
\cdot
\begin{bmatrix}
E_{0}\\
E_{1}\\
E_{2}\\
E_{3}\\
E_{4}\\
E_{5}
\end{bmatrix}
= 0
\end{equation}
They are evaluated at $\rho = \rho_{0}$ where $\rho_{0}$ is the radius of the inner-most bin $r=0$.
The ``intermediate'' relations are:
\begin{equation}
\small
\begin{bmatrix}
-W_{ (4r-2) \theta} & -W_{(4r-1)\theta} &-W_{(4r)\theta} & -W_{(4r+1)\theta}& W_{(4r+2)\theta} &W_{(4r+3)\theta} & W_{(4r+4)\theta} &W_{(4r+5)\theta}\\
-M_{(4r-2)} & -M_{(4r-1)}& -M_{(4r)} & -M_{(4r+1)} &M_{(4r+2)} &M_{(4r+3)}&M_{(4r+4)} &M_{(4r+5)}\\
-\mathcal{D}_{(4r-2)} & -\mathcal{D}_{(4r-1)}& -\mathcal{D}_{(4r)} & -\mathcal{D}_{(4r+1)} &\mathcal{D}_{(4r+2)} &\mathcal{D}_{(4r+3)}&\mathcal{D}_{(4r+4)} &\mathcal{D}_{(4r+5)}\\
-W_{ (4r-2) z} & -W_{(4r-1)z} &-W_{(4r)z} & -W_{(4r+1)z}& W_{(4r+2)z} &W_{(4r+3)z}& W_{(4r+4)z} &W_{(4r+5)z}
\end{bmatrix}
\cdot
\begin{bmatrix}
E_{(4r-2)}\\
E_{(4r-1)}\\
E_{(4r)}\\
E_{(4r+1)}\\
E_{(4r+2)}\\
E_{(4r+3)}\\
E_{(4r+4)}\\
E_{(4r+5)}
\end{bmatrix}
= 0
\end{equation}
and are evaluated at $\rho = \rho_{r}$ for $0<r<R$.
The outer-most relations are:
\begin{equation}
\small
\begin{bmatrix}
-W_{ (4R-2) \theta} & -W_{(4R-1)\theta} &-W_{(4R)\theta} & -W_{(4R+1)\theta}& W_{(4R+2)\theta} &W_{(4R+3)\theta}\\
-M_{(4R-2)} & -M_{(4R-1)}& -M_{(4R)} & -M_{(4R+1)} &M_{(4R+2)} &M_{(4R+3)}\\
-\mathcal{D}_{(4R-2)} & -\mathcal{D}_{(4R-1)}& -\mathcal{D}_{(4R)} & -\mathcal{D}_{(4R+1)} &\mathcal{D}_{(4R+2)} &\mathcal{D}_{(4R+3)}\\
-W_{ (4R-2) z} & -W_{(4R-1)z} &-W_{(4R)z} & -W_{(4R+1)z}& W_{(4R+2)z} &W_{(4R+3)z}
\end{bmatrix}
\cdot
\begin{bmatrix}
E_{(4R-2)}\\
E_{(4R-1)}\\
E_{(4R)}\\
E_{(4R+1)}\\
E_{(4R+2)}\\
E_{(4R+3)}
\end{bmatrix}
= -E_{ (4R+4) }
\begin{bmatrix}
W_{(4R+4)\theta}\\
M_{(4R+4)}\\
\mathcal{D}_{(4R+4)}\\
W_{(4R+4)z}
\end{bmatrix}
\end{equation}
evaluated at $\rho = \rho_{R} = a_b$.

\subsection{Appendix D: Derivation of scattering-width}
\label{sec:A4}
First, derive the ratio of the power scattered to the power incident: $\frac{P_{sct}}{P_{inc}}$. The power is $P = \int d\bold{a}\cdot \bold{S}$, where $\bold{S}$ is the time-averaged Poynting flux and $\bold{a}$ is the cross-sectional area of interest. Consider the incident power through the cross-sectional area of dimensions $L_z$ and $L_y$ on the yz- plane.
\begin{equation}
P_{inc,x} = S_{inc,x}L_y L_z
\end{equation}
The equation above is straight-forward since $S_{inc,x}$ is assumed constant. The scattered power radiating away from cylinder is:
\begin{equation}
P_{sct,\rho}(\rho) = \rho L_z \int_{-\pi}^{+\pi} S_{sct,\rho}(\rho,\theta)d\theta
\end{equation}
Only the far-field radiation is considered (and therefore multi-pole effects near the cylinder are neglected). In this case:
\begin{equation}
P_{sct,\rho}|_{\textrm{far-field}} = \textrm{lim}_{\rho \rightarrow \infty} \rho L_z \int_{-\pi}^{+\pi} S_{sct,\rho}(\rho,\theta)d\theta 
\end{equation}
In general, the far-field radial scattered power $P_{sct,\rho}|_{\textrm{far-field}}$ converges to a non-zero value because $S_{sct,\rho}(\rho,\theta)\propto 1/\rho$ for large $\rho$. From now on, $P_{sct,\rho}$ is taken to mean $P_{sct,\rho}|_{\textrm{far-field}}$. Next, define the scattering-width, $\sigma$:
\begin{equation}
\sigma \equiv \frac{P_{sct} L_{y}}{P_{inc}} = \frac{\textrm{lim}_{\rho \rightarrow \infty} \rho \int_{-\pi}^{+\pi} S_{sct,\rho}(\rho,\theta)d\theta }{S_{inc,x}}
\end{equation}
which has the physical meaning of power scattered per cylinder \emph{per} incident power/$L_{y}$ \cite{myra2010scattering}. Clearly, as $L_y$ (the incident beam-width in the y-direction) increases, less power is directly incident on the cylinder. So as $L_{y} \rightarrow \infty$, also $\frac{P_{sct}}{P_{inc}} \rightarrow 0$. In reality, the Lower-Hybrid beam has a finite width, and there are multiple cylinders (SOL filaments) in its path. This allows the cancellation of the $L_{y}$ variable. Suppose a beam of width $L_y$ is traveling through a turbulent layer of width $L_x$. Within that layer are filaments of average radius $a_b$. Assuming the cross-sectional packing fraction $f_{p}$ of filaments in this layer is known, the beam encounters $\frac{f_{p} L_x L_y}{\pi a_b^2}$ filaments on average. This can be used to roughly estimate the fraction of incident power scattered from multiple filaments:
\begin{equation}
\frac{P_{sct}}{P_{inc}} \approx \frac{f_{p} L_x}{\pi a_b^2} \sigma
\end{equation}
This is only valid for a \emph{sparse} filament layer, because the effects of an already scattered wave interacting with another filament are ignored. This is properly accounted for in the RTE introduced in \hyperref[sec:5]{Section 5}.

\end{appendix}

\section*{Acknowledgements}
This work was supported by US DoE under contract numbers: DE-SC0018090 supporting the RF-SciDAC 4 project, and DE-SC0014264 supporting PSFC MFE projects. The GENRAY/CQL3D and PETRA-M simulations presented in this paper were performed on the MIT-PSFC partition of the Engaging cluster at the MGHPCC facility (www.mghpcc.org) which was funded by DoE grant number DE-FG02-91-ER54109.

\section*{Declaration of interest}
The authors report no conflict of interest.

\newpage

\bibliography{SAS_MC_paper_arxiv}
\end{document}